\documentclass[aps,prd,nofootinbib,showpacs,superscriptaddress,twocolumn,floatfix]{revtex4-1}
\usepackage{amsmath,amssymb,bm,color,fancyhdr,hyperref,lastpage,units}
\usepackage{float}
\usepackage[latin1,utf8]{inputenc}
\usepackage[english]{babel}

\usepackage{slashed}
\usepackage{csquotes}
\usepackage{comment}
\usepackage{graphicx}
\usepackage{dcolumn}
\usepackage{bm}
\usepackage{appendix}
\usepackage{hyperref}
\usepackage[normalem]{ulem}

\newcommand{\pbp}{\langle \bar{\psi}\psi \rangle}

\begin{document}

\title{Chiral condensates and screening masses of neutral pseudoscalar mesons \\
from lattice QCD at physical quark masses}

\author{Heng-Tong Ding}
\affiliation{Key Laboratory of Quark and Lepton Physics (MOE) and Institute of
Particle Physics, Central China Normal University, Wuhan 430079, China}

\author{Jin-Biao Gu}
\affiliation{Key Laboratory of Quark and Lepton Physics (MOE) and Institute of
Particle Physics, Central China Normal University, Wuhan 430079, China}

\author{Sheng-Tai Li}
\affiliation{Key Laboratory of Quark and Lepton Physics (MOE) and Institute of
Particle Physics, Central China Normal University, Wuhan 430079, China}

\author{Rishabh Thakkar}
\affiliation{Key Laboratory of Quark and Lepton Physics (MOE) and Institute of
Particle Physics, Central China Normal University, Wuhan 430079, China}
\date{\today}

\begin{abstract}
We investigate the effects of temperature $T$ and external magnetic fields $eB$ on the chiral condensates and screening masses of neutral pseudoscalar mesons, including $\pi^0$, $K^0$, and $\eta_{s\bar{s}}^0$, in (2+1)-flavor lattice QCD with physical quark masses. The chiral condensates are intrinsically connected to the screening masses via Ward-Takahashi identities, with the latter characterizing the inverse of the spatial correlation length in the pseudoscalar channel. Using highly improved staggered quarks, we perform simulations on lattices with temporal extents $N_\tau = 8, 12, 16$ and an aspect ratio of 4, covering five temperatures from 145 MeV to 166 MeV. For each temperature, eight magnetic field strengths are simulated, reaching up to $eB \sim 0.8$ GeV$^2$. These simulations allow us to provide continuum estimates for the chiral condensates and screening masses. We observe intricate behavior in the light ($ud$), strange-light ($ds$) and strange ($s$) quark condensates as functions of the magnetic field and temperature, reflecting the competition between magnetic catalysis and inverse magnetic catalysis effects. 
This complex behavior is also mirrored in the screening masses of the neutral pseudoscalar mesons. Notably, the screening masses of $\pi^0$ and $K^0$ exhibit a non-monotonic dependence on $eB$, closely following the variations in their corresponding chiral condensates. Meanwhile, the screening mass of $\eta_{s\bar{s}}^0$ decreases monotonically with increasing $eB$. These findings provide valuable insights for understanding the behavior of QCD in a thermomagnetic medium and can serve as benchmarks for low-energy QCD models and effective theories.
\end{abstract}

\maketitle
\section{INTRODUCTION}

The influence of magnetic fields on the QCD thermal medium is crucial for understanding the behavior of physical systems where the magnetic field strength approaches the QCD scale, $\sqrt{eB} \sim \Lambda_\mathrm{QCD}$. Magnetic fields with strengths $\sqrt{eB} \sim 2$ GeV are hypothesized to have existed in the early universe during the electroweak phase transition~\cite{Vachaspati:1991nm,Enqvist:1993np,Baym:1995fk,Grasso:2000wj}. Similarly strong magnetic fields are generated in heavy-ion collision experiments, reaching magnitudes of approximately $\sqrt{eB} \sim 0.1$ GeV at the Relativistic Heavy Ion Collider (RHIC) and up to $\sqrt{eB} \sim 0.5$ GeV at the Large Hadron Collider (LHC)~\cite{Kharzeev:2020jxw,Skokov:2009qp,Voronyuk:2011jd,Deng:2012pc,Kharzeev:2020jxw}. Additionally, comparable magnetic field strengths, around $\sqrt{eB} \sim 1$ MeV, are conjectured to occur in astrophysical systems, particularly in certain neutron stars known as magnetars~\cite{Duncan:1992hi}. Understanding the interplay between magnetic fields and the strong force is therefore essential for exploring these diverse physical systems.

Strong magnetic fields significantly affect QCD observables, offering insights into deconfinement and chiral symmetry restoration (see~\cite{Andersen:2014xxa,Cao:2021rwx,Endrodi:2024cqn,Adhikari:2024bfa} for recent reviews). For instance, conserved charge fluctuations~\cite{Ding:2021cwv,Ding:2020pao,Ding:2023bft}, Polyakov loops~\cite{Bruckmann:2013oba,DElia:2018xwo}, string tension~\cite{Bonati:2016kxj}, and the ratio of pressure to energy density~\cite{Bali:2014kia} probe the deconfinement aspect of the transition. In terms of the chiral aspect, magnetic fields alter the chiral symmetry from $SU_L(2) \times SU_R(2) \sim O(4)$ to $U_L(1) \times U_R(1) \sim O(2)$ in the case of two-flavor QCD. In the absence of a magnetic field, two-flavor QCD features three degenerate Goldstone pions arising from spontaneous chiral symmetry breaking. However, with a nonzero magnetic field, only the neutral pion remains as the Goldstone boson, while the charged pions become heavier due to explicit symmetry breaking by the magnetic field.

A key phenomenon in this context is inverse magnetic catalysis, where the light-quark chiral condensate, an order parameter for chiral symmetry restoration, decreases with increasing magnetic field strength near the transition temperature~\cite{Bali:2012zg,Bali:2011qj,Bornyakov:2013eya,Tomiya:2019nym,DElia:2018xwo,Endrodi:2019zrl,Ding:2022tqn}. This suppression of chiral condensate and the accompanied reduction of the pseudocritical temperature $T_{pc}$ are not predicted by effective theories~\cite{Shovkovy:2012zn}, making inverse magnetic catalysis a hallmark effect observed in lattice studies. In contrast, at zero temperature, magnetic catalysis occurs, where the chiral condensate increases with $eB$~\cite{Shovkovy:2012zn,Endrodi:2024cqn}.

It has been pointed out in Ref. \cite{Ding:2022tqn} by some of the current authors that the chiral condensates at nonzero temperature and magnetic field are intrinsically linked to the corresponding space-time integral of the two-point correlation functions in the pseudoscalar channel. The exponential decay of these two-point correlation functions defines the screening mass, which characterizes the long-distance behavior of the system and is inversely related to the screening length. As a result, the inverse of the screening mass exhibits a dependence on both the temperature and the magnetic field strength similar to that of the chiral condensate. This connection highlights the intimate relationship between chiral symmetry restoration and the screening properties of the QCD medium, providing a unified picture of the system's response to external conditions. 

However, results presented in Ref. \cite{Ding:2022tqn} were obtained from lattice QCD simulations on $32^3 \times N_\tau$ lattices at one single lattice cutoff ($a = 0.117 \, \text{fm}$) using a fixed-scale approach with a larger-than-physical pion mass of approximately 220 MeV \cite{Ding:2022tqn}. While chiral condensates and meson screening masses in external magnetic fields have been extensively studied using effective theories and models~\cite{Coppola:2024uvz,Wen:2024hgu,Mao:2024rxh,Ayala:2023llp,Sheng:2021evj,Andersen:2014xxa,Cao:2021rwx,Endrodi:2024cqn,Adhikari:2024bfa,Moreira:2022dwo}, achieving continuum-extrapolated results from lattice QCD simulations with physical pion masses remains essential. Such results are crucial for obtaining a more realistic and quantitative understanding of these phenomena and for meaningful comparisons with studies based on effective theories and models.

In this study, we significantly extend our previous work \cite{Ding:2022tqn} by performing lattice simulations with a physical pion mass on $N_\tau = 8, 12,$ and 16 lattices, maintaining an aspect ratio of 4. These simulations cover five different temperature values around the pseudo-transition temperature and eight values of the external magnetic field strength ($eB$) at each temperature. This setup allows for a continuum estimate of the up, down, and strange quark chiral condensates, as well as the screening masses of neutral mesons in the pseudoscalar channel.

The structure of the paper is organized as follows: In Sec. \ref{theory}, we introduce the theoretical framework, focusing on the Ward-Takahashi identity that connects the chiral condensates and meson susceptibilities, along with essential observables such as the chiral condensates, two-point correlation functions for mesons, and screening masses. In Sec. \ref{lattice}, we describe the setup for our lattice simulations using  QCD with highly improved staggered fermions as well as the lattice observables and methodologies. Sec. \ref{results} presents the primary findings containing the continuum limit $T-eB$ dependence of chiral condensate and mesonic screening mass. Lastly, Sec. \ref{conclusion} summarizes our conclusions. We also provide details regarding the interpolation across the $T$-$eB$ plane and continuum estimate in Appendix \ref{interpolation}, continuum estimates of renormalized chiral condensates separately for the $u$ and $d$ quarks and their $T$ and $eB$ dependence in Appendix \ref{udquark} and outline the statistics of the lattices used in the study in Appendix \ref{statistics}.

\section{chiral condensates, susceptibilities and screening masses}
\label{theory}
Both the quark chiral condensates and the neutral pseudoscalar meson correlation function offer distinct insights into the QCD thermomagnetic medium. Notably, they are intricately linked through the Ward-Takahashi identities at nonzero magnetic fields \cite{Ding:2020hxw}:
\begin{eqnarray}
(m_u+m_d)\,\chi_{\pi^0}&=&\pbp_u+\pbp_d\,,\label{eq:Ward-id_pi}\\
(m_d+m_s)\,\chi_{K^0}&=&\pbp_d+\pbp_s\,,\\
m_s\,\chi_{\eta^0_{s\bar{s}}}&=&\pbp_s\,.
\label{eq:Ward-id}
\end{eqnarray}
Here, $\pbp_f$ and $m_f$ represent the chiral condensate and quark mass for up, down and strange quark mass denoted by $u,d$ and $s$ respectively. $\chi_H$ denote the neutral pseudoscalar meson susceptibility, defined as the integral of the Euclidean two-point correlation function for three neutral pseudoscalar mesons $H=\pi^0, K^0$ and $\eta^0_{s\bar{s}}$,
\begin{eqnarray}
    \chi_H(B,T)&\equiv&\int\text{d}z\int_{0}^{1/T}\text{d}\tau\int\text{d}x\int\text{d}y\,\mathcal{G}_H(B,\boldsymbol{x})\nonumber\\
    &=&\int\text{d}z\,G_H(B,T,z)\,,
    \label{susceptibility_def}
\end{eqnarray}
where $\boldsymbol{x}\equiv(\tau,\overrightarrow{x})=(\tau,x,y,z)$, $B$ is the external magnetic field strength, $T$ is the temperature, and the meson correlation function $\mathcal{G}_{f_1,f_2}(B,\boldsymbol{x})=\mathcal{O}_{f_1,f_2}(B,\boldsymbol{x})\mathcal{O}^\dagger_{f_1,f_2}(B,0)$ is obtained from the meson interpolator $\mathcal{O}_{f_1,f_2}(B,\boldsymbol{x})=\bar{\psi}_{f_1}(B,\boldsymbol{x})\gamma_5\psi_{f_2}(B,\boldsymbol{x})$. The meson spatial correlator $G_H(B, T,z)$ is derived from the meson correlation function by projecting over zero momentum in the $x,y$ and $\tau$ directions. Thus, using Ward-Takahashi identities we can probe the nexus between the magnetic field dependence of the chiral condensate and the properties of the Goldstone mesons, $\pi^0$ and $K^0$, as well as the fictitious lightest neutral pseudoscalar from strange quarks $\eta^0_{s\bar{s}}$ through their correlation functions. 

The quark condensate $\pbp_f(B,T)$ is obtained from the QCD partition function $\mathcal{Z}$ by taking a derivative with respective quark mass 
\begin{equation}
    \pbp_f(B,T)=\frac{1}{N_\sigma^3N_\tau} \frac{\partial\text{ln}\mathcal{Z}(B,T)}{\partial m_f}=\frac{1}{N_\sigma^3N_\tau}\text{Tr} M_f^{-1}\,,
\end{equation}
where the QCD partition function is defined as 
\begin{equation}
    \mathcal{Z}(B,T)=\int \mathcal{D}U\,e^{-S_g}\prod_{f=u,d,s}\text{det}M_f \,,
\end{equation}
and $U$ is the gauge field.

In order to facilitate a continuum limit and to observe the effect of magnetic effect on the chiral condensate, we get rid of the additive and multiplicative divergences using the following combination \cite{Bali:2012zg,Ding:2022tqn}:
\begin{eqnarray}
    \Delta\Sigma_{ud}(B,T) &= \frac{m_u+m_d}{2M_\pi^2f_{\pi}^2}&\sum_{f=u,d} \left\{\pbp_f(B,T) \right. \nonumber\\
    && \left. - \pbp_f(0,T)\right\}, \\
    \Delta\Sigma_{ds}(B,T) &= \frac{m_d+m_s}{2M_K^2f_K^2}&\sum_{f=d,s} \left\{\pbp_f(B,T) \right. \nonumber\\
    && \left. - \pbp_f(0,T)\right\}, \\
    \Delta\Sigma_{s}(B,T) &= \frac{m_d+m_s}{2M_K^2f_K^2}&\left\{ \pbp_s(B,T) \right. \nonumber\\
    && \left. - \pbp_s(0,T)\right\}\,,
\end{eqnarray}
where $M_H$ and $f_H$ are the meson mass and meson decay constant for mesons $H=\pi$ and $K$ at vanishing magnetic fields. In this study, following~\cite{Bali:2012zg} we adopt the chiral limit of the decay constants for our normalization factors, using $f_\pi=86$ MeV and $f_K/f_\pi=1.2$ ~\cite{Colangelo:2003hf,Colangelo:2005gd} and physical mass values for pion $M_\pi=135$ MeV and kaon $M_K=498$ MeV. We got rid of the additive divergences by subtracting the corresponding chiral condensate at zero magnetic fields and the multiplicative divergence by multiplying the quark mass. The observables are normalized by $M_H^2f_H^2$ to make the renormalized quantity dimensionless, and the combination is obtained from three flavors Gell-Mann-Oakes-Renner (GMOR) relation \cite{Ding:2020hxw,GellMann:1968rz,Gasser:1984gg}.

The screening correlators $G_H(B, T, z)$, defined earlier, decay exponentially at large spatial distances \cite{Detar:1987kae, Detar:1987hib}:  
\begin{eqnarray}
    \underset{z \to \infty}{\lim} G_H(B, T, z) = A_H e^{-M_H z},
\end{eqnarray}  
where $A_H$ and $M_H$ represent the amplitude and the screening mass for hadron $H$, respectively. At shorter distances, contributions from excited states become significant. On the lattice, due to its periodic boundary conditions and finite spatial extent, the screening correlator exhibits symmetry around $N_\sigma/2$. Furthermore, for staggered mesons, the correlator typically oscillates, as it couples two sets of mesons with the same spin and opposite parities. This behavior can be captured using the following ansatz \cite{Bazavov:2019www, Cheng:2010fe}:  
\begin{eqnarray}
    G_H(B, T, n_z) = \sum_{i} A_H^{i, nosc} \cosh \big[M_H^{i, nosc}(N_\sigma/2 - n_z)\big] \nonumber\\
    - (-1)^{n_z} \sum_{j} A_H^{j, osc} \cosh \big[M_H^{j, osc}(N_\sigma/2 - n_z)\big],\nonumber\\
\end{eqnarray}  
where the indices $nosc$ and $osc$ denote the non-oscillating and oscillating contributions, respectively, and $n_z$ represents the distance in the $z$ direction in lattice units.  

For the neutral pseudoscalar meson, the parity partner corresponds to the conserved charge, which does not excite any states. However, the presence of a magnetic field induces mixing between the neutral pion and the $\rho$-meson with $s_z = 0$ \cite{Bali:2017ian, Ding:2020hxw}. This mixing modifies the correlator structure and, thus, requires the need to use oscillating correlators.

The screening mass, $M_H$, provides insight into the medium's long-distance behavior and is inversely related to the correlation length for hadron $H$. At zero temperature, the screening mass reduces to the pole mass of the hadron's ground state \cite{Karsch:2003jg}. Unlike observables such as susceptibility, which predominantly capture short-distance effects, the screening mass offers a window into the medium's behavior over extended spatial scales, making it a critical observable for studying the interplay of temperature, magnetic fields, and QCD medium properties.

\section{LATTICE SIMULATION}
\label{lattice}
Our current simulations have been performed for $N_f = 2 + 1$ QCD with highly improved staggered quark (HISQ) fermions \cite{Follana:2006rc} and a tree-level improved Symanzik gauge action, similar to the ones used by the HotQCD Collaboration \cite{Bazavov:2011nk,Bazavov:2012jq,Bazavov:2014pvz,Bazavov:2017dus,Bazavov:2018mes,Bollweg:2024epj}. The up and down light quarks are assumed to have identical masses with distinct behavior under a magnetic field due to different electric charges. The magnetic field has been introduced along the $z$ direction by multiplying the gauge field with the complex phase factor. The finite volume with periodic boundary conditions of lattice quantizes the magnetic flux leading to the following quantization condition \cite{Bali:2011qj,DElia:2010abb}
\begin{eqnarray}
    eB=\frac{6\pi N_b}{N_\sigma^2a^2}=6\pi N_bT^2\frac{N_\tau^2}{N_\sigma^2}\,,
\end{eqnarray}
where $a$ is the lattice spacing and $N_b\in\boldsymbol{Z}$ represents the magnetic flux through a unit area in a plane perpendicular to $z$ direction. Further details about the magnetic field quantization and its implementation on the lattice can be found in \cite{Ding:2020hxw}.

For our analysis, to perform the continuum limit, we have fixed the volume of the system by fixing the aspect ratio of the lattices $N_\sigma/N_\tau=4$ and considered the lattice sized with $N_\tau=8,12$ and $16$. The strange quark mass $m_s$ was adjusted to its physical value through the procedure outlined in \cite{Ding:2020hxw} by tuning the mass of the fictitious $s\bar{s}$ pseudoscalar meson, $\eta^0_{s\bar{s}}$ to $M_{\eta^0_{s\bar{s}}}=\sqrt{2M_K^2-M_\pi^2}\simeq 684$ MeV \cite{Bazavov:2014cta,Bazavov:2019www}. The mass of $\eta^0_{s\bar{s}}$ is estimated using the leading order chiral perturbation theory calculation. The light quark mass $m_l$ was fixed at $m_l=m_s/27$ giving the pion mass $M_\pi\simeq135$ MeV. The scale was set using the kaon decay constant $f_Ka(\beta)$ using the parametrization obtained in \cite{Bazavov:2019www}. For each lattice size, we investigated five temperatures ranging from 145 MeV to 165 MeV, in proximity to the critical temperature, with eight $N_b$ values between 0 to 24 for each temperature corresponding to magnetic field strengths ($eB$) up to approximately 0.8 GeV$^2$. All gauge configurations were generated using a modified version of SIMULATeQCD \cite{HotQCD:2023ghu}, with data saved at every tenth trajectory. The statistics are listed in Table~\autoref{tab:328_new}, ~\autoref{tab:4812_new}, and~\autoref{tab:6416_new} in Appendix~\ref{statistics}.

The chiral condensates are computed on-the-fly during the generation of gauge configurations. Their analysis is performed every 10th trajectory using 10 random Gaussian sources per configuration. The two-point correlation functions are obtained on each gauge configuration using corner wall sources. The contribution of the disconnected quark lines to the screening mass was neglected in this analysis, as their impact is expected to be small~\cite{Luschevskaya:2015cko,Ding:2020hxw}. Single corner wall sources were used for lattices with $N_\tau=8$ and 12 while four corner wall sources were used for lattices with $N_\tau=16$ to reduce noise. To obtain the lattice screening mass, we used multiple fits (up to 3 states) and selected using the corrected Akaike information criteria (AICc) using a procedure similar to the ones used in \cite{Bazavov:2019www,Ding:2020hxw,Ding:2022tqn}. The procedures for obtaining the continuum estimate along with intermediate $T$ and $eB$ for each lattice size are outlined in Appendix \ref{interpolation}.

\section{RESULTS}
\label{results}
\begin{figure}
\includegraphics[scale=0.63]{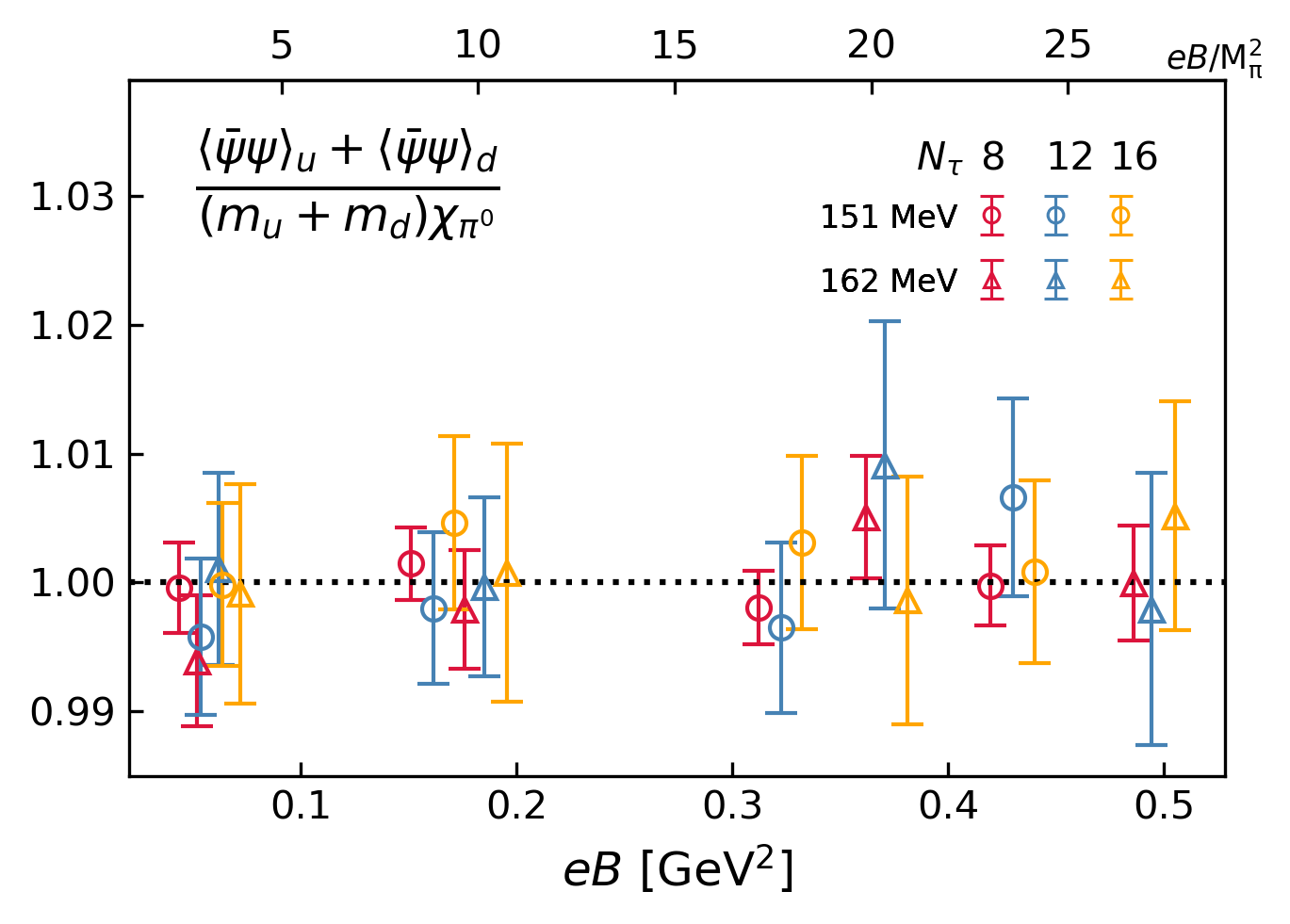}
\includegraphics[scale=0.63]{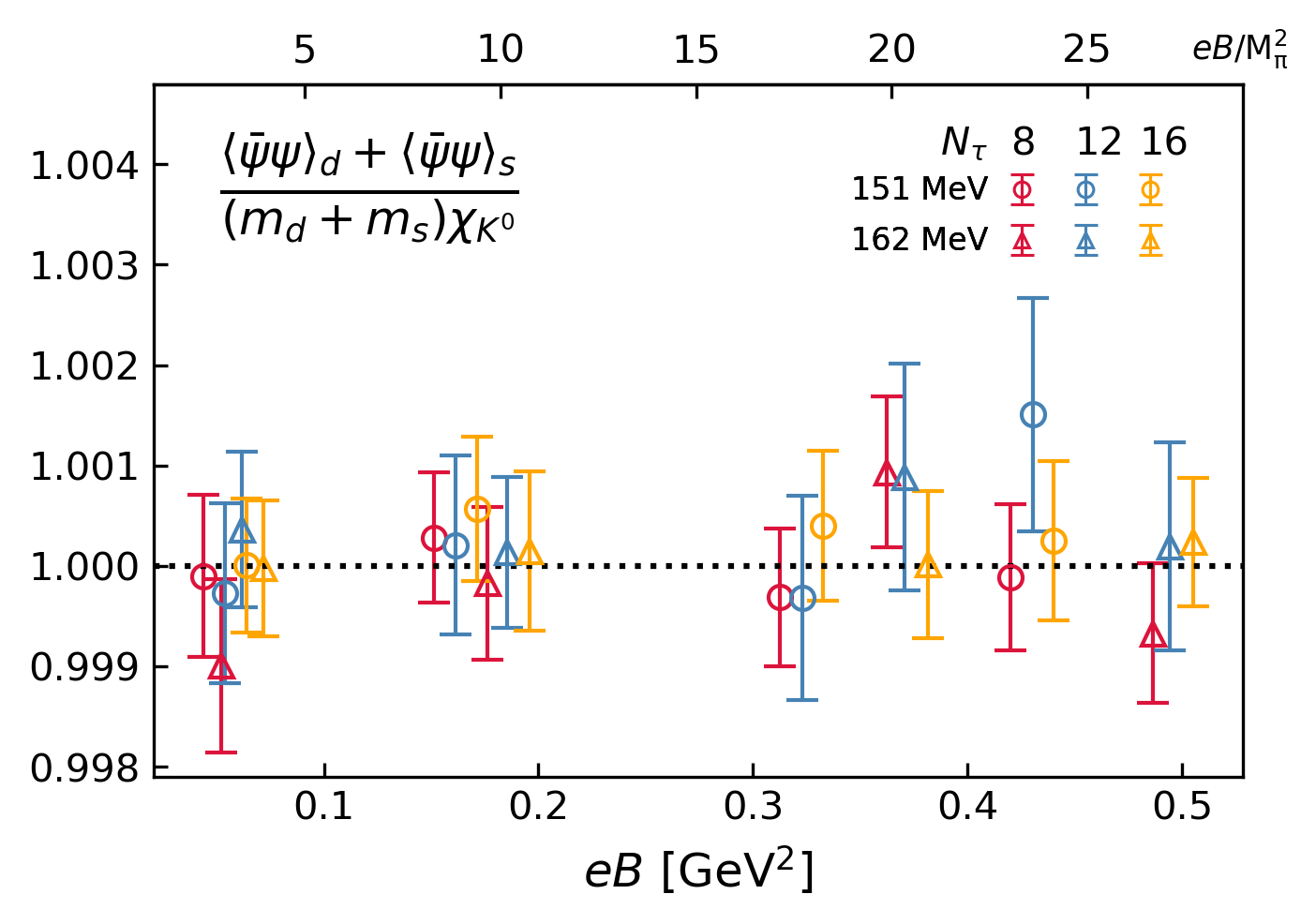}
\includegraphics[scale=0.63]{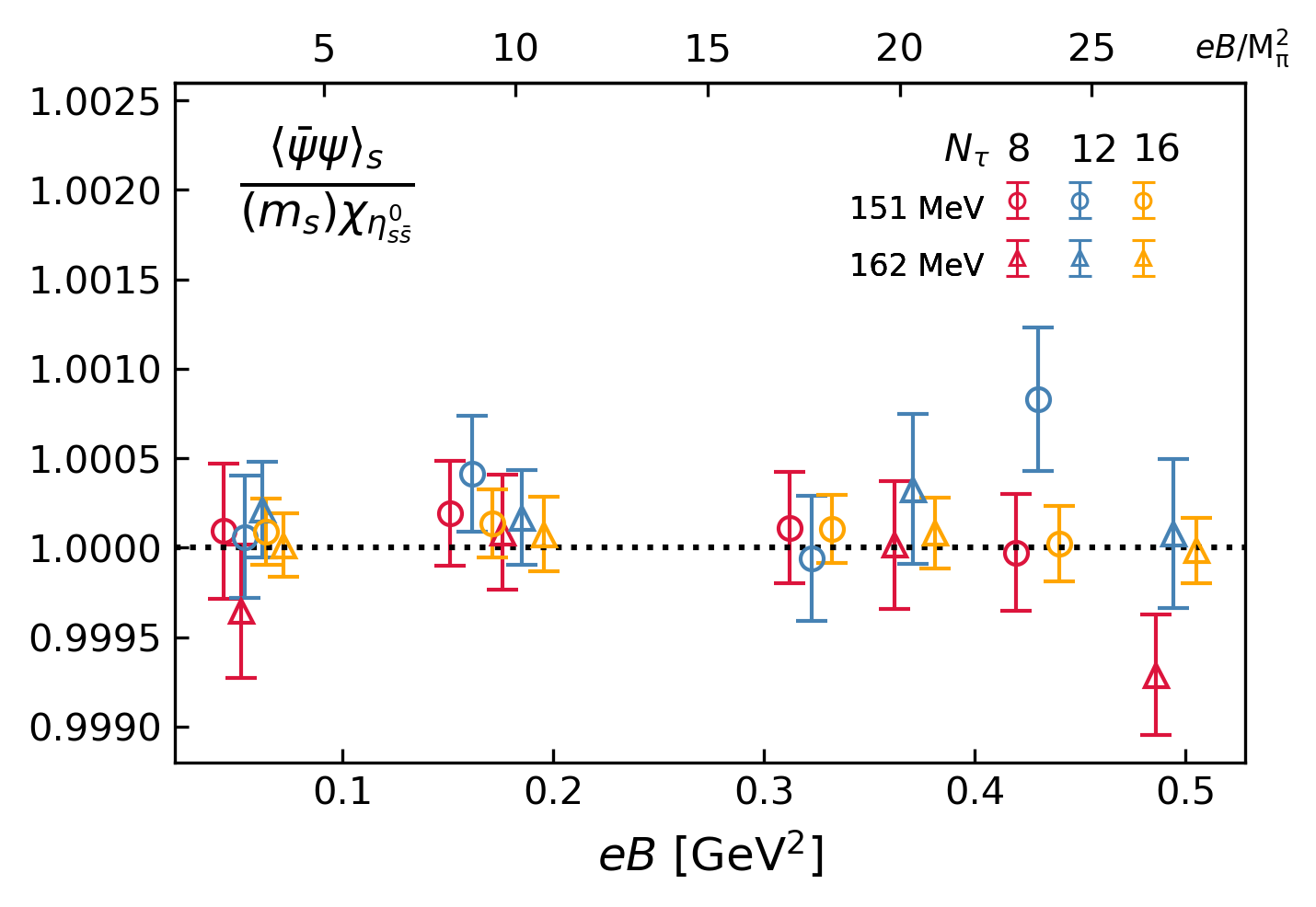}
\caption{Combination of chiral condensates normalized with respect to their corresponding quark masses and susceptibility for the $\pi^0$ (top), $K^0$ (middle), and $\eta^0_{s\bar{s}}$ (bottom) mesons, respectively, as a function of the magnetic field strength $eB$ for three lattice sizes with temporal extent $N_\tau=8,12$ and 16 at two temperatures (rounded to nearest integer) $T$ = 151 MeV and 162 MeV. Data points for $N_\tau = 8$ and $N_\tau = 16$ have been shifted left and right by 0.01 GeV$^2$, respectively, to enhance visibility. The upper $x$-axis is rescaled by the pion mass square in the vacuum at $eB = 0$ to make it dimensionless. }
\label{wi}
\end{figure}
\begin{figure}
\includegraphics[scale=0.63]{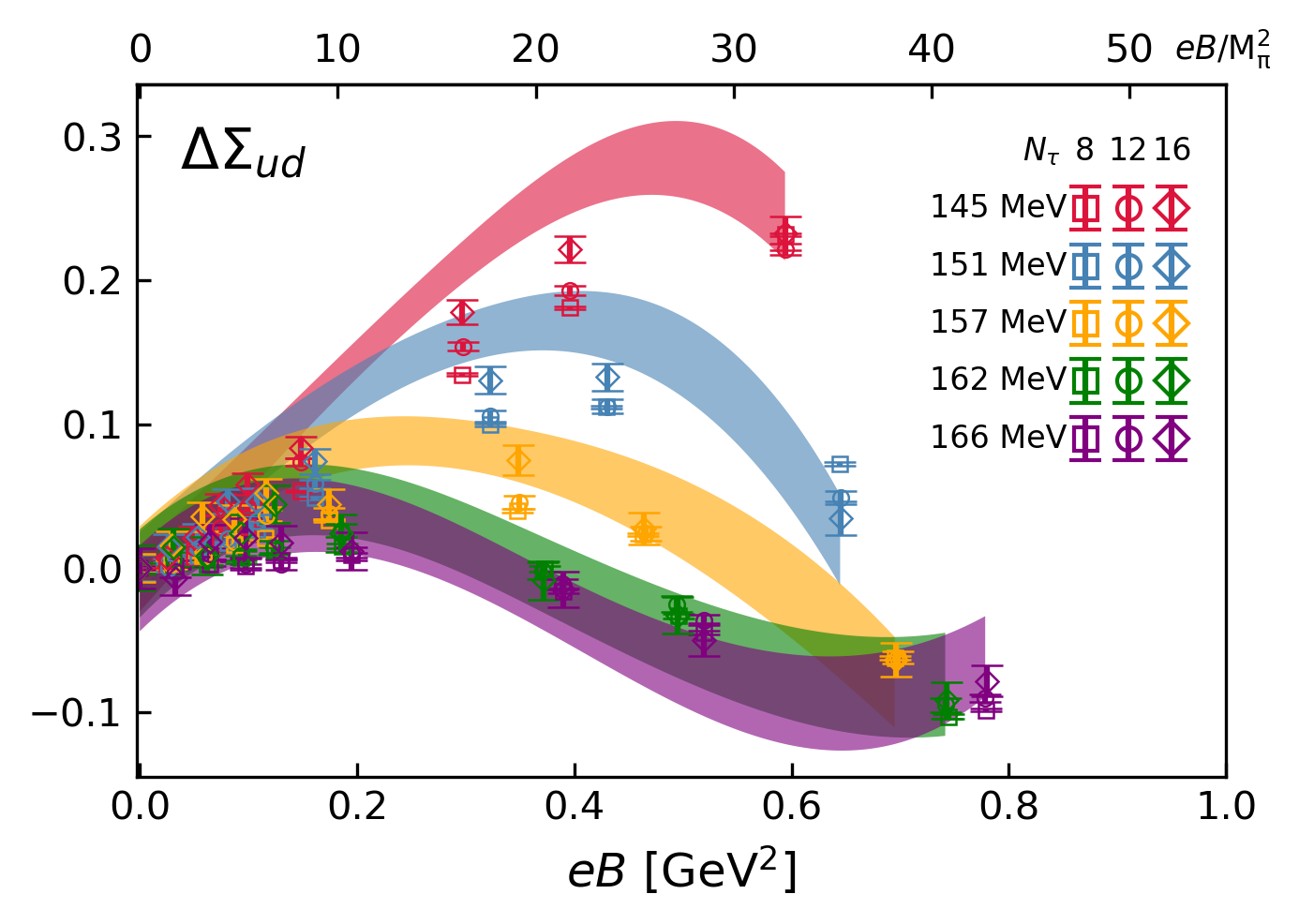}
\includegraphics[scale=0.63]{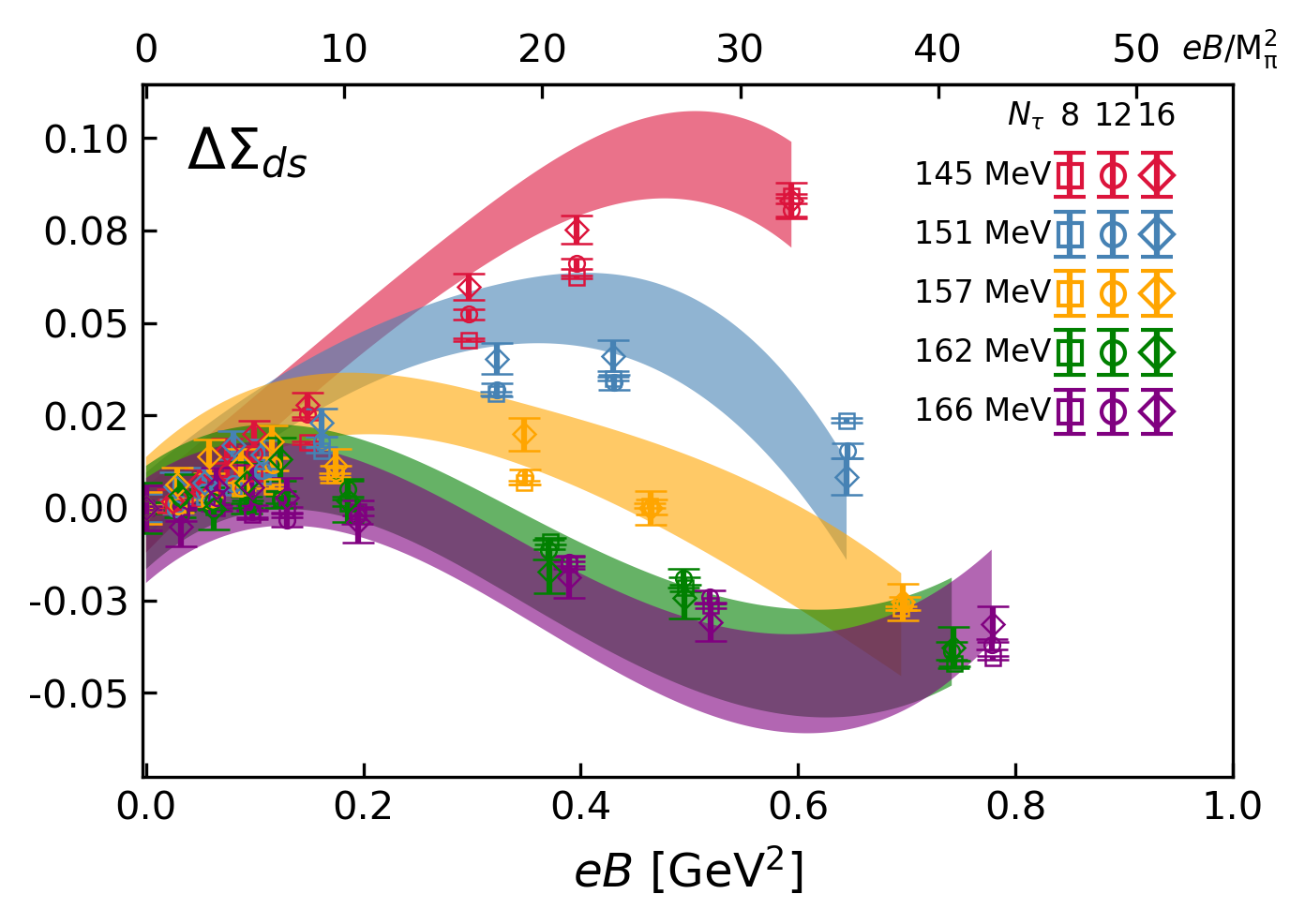}
\includegraphics[scale=0.63]{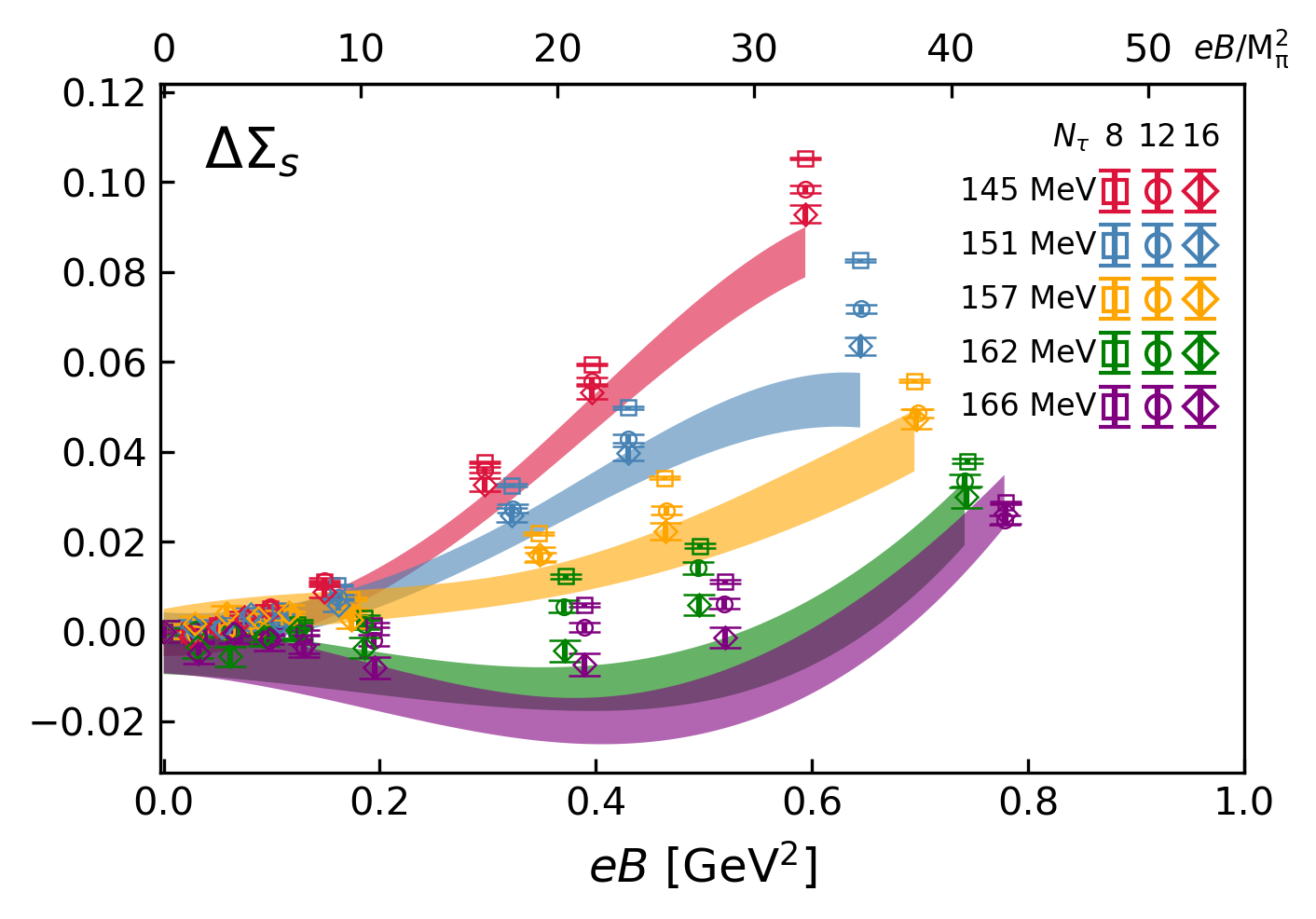}
\caption{The continuum estimate of the change of the renormalized 
chiral condensates $\Delta\Sigma_{ud}$ (top), $\Delta\Sigma_{ds}$ (middle), $\Delta\Sigma_{s}$ (bottom) as a function of the magnetic field strength
$eB$ at fixed temperatures. The shaded bands represent the continuum estimated results, while the data points correspond to lattice data at temperatures rounded to the nearest integer (refer to Appendix \ref{statistics} for exact temperature values for each lattice data). The upper $x$-axis is rescaled by the pion mass square in the vacuum at $eB = 0$ to make it dimensionless. 
}
\label{pbp_vs_eb}
\end{figure}
As discussed in Section \ref{theory}, the Ward-Takahashi identities play a crucial role in establishing the connection between the chiral condensate and the correlation functions.

Although these identities hold exactly in the continuum limit, discretization effects inherent to staggered fermions on the lattice could lead to small deviations at finite lattice spacing. Moreover, in our numerical verification, we have neglected contributions from the disconnected diagrams in the neutral pion ($\pi^0$) correlation functions, which could further affect numerical accuracy.
In \autoref{wi}, we verify the validity of these identities using our lattice data. The figure presents the ratios of chiral condensates, normalized by their corresponding quark masses and susceptibilities, for the $\pi^0$ (top), $K^0$ (middle) and $\eta^0_{s\bar{s}}$ (bottom) mesons across three lattice sizes and at two temperatures ($T$ = 151 MeV and 162 MeV, rounded to the nearest integer). Although small deviations due to lattice discretization effects and the neglected disconnected contributions in the neutral pion correlation functions might be present, our lattice data
still consistently cluster around unity, supporting the robustness of the Ward-Takahashi identities at the lattice spacings studied.

The impact of the magnetic field on the chiral condensate is illustrated in \autoref{pbp_vs_eb}. The figure presents the variation of the renormalized chiral condensate as a function of the magnetic field strength $eB$ at fixed temperature values for three different quark combinations: $\Delta\Sigma_{ud}$ (top), $\Delta\Sigma_{ds}$ (middle) and $\Delta\Sigma_{s}$ (bottom) as defined in Section \ref{theory}. The shaded bands in the figure depict the continuum estimated results, while the data points represent lattice measurements at temperatures rounded to the nearest integer~\footnote{The exact temperature values for each lattice data point are provided in Appendix~\ref{statistics}.}. The continuum estimate of observables in our studies is obtained using a linear and quadratic ansatz in $1/N_\tau^2$, based on results from simulations on $N_\tau = 8$, 12, and 16 lattices. Details of the continuum estimation can be found in Appendix~\ref{interpolation}. Our continuum estimated results for $\Delta\Sigma_{ud}$, exhibit a similar trend to those reported in previous analysis \cite{Bali:2012zg}, where $N_\tau=6$, 8 and 10 were used with a linear ansatz in $1/N_\tau^2$ to obtain the continuum limit, though some deviations beyond the reported uncertainties are observed.

Both $\Delta\Sigma_{ud}$ and $\Delta\Sigma_{ds}$ exhibit similar trends as the magnetic field strength increases. At lower temperatures, the renormalized chiral condensate initially increases with $eB$, reaching a peak before decreasing in magnitude within the available range of $eB$. This behavior indicates the dominance of magnetic catalysis at smaller $eB$, followed by the onset of inverse magnetic catalysis at larger $eB$. As the temperature increases, the position of this peak shifts to lower values of $eB$. Specifically, the peak position changes from $\sim$ 0.46 GeV$^2$ at the lowest temperature to $\sim$ 0.1 GeV$^2$ at the highest temperature. Additionally, at two highest temperatures, after reaching the peak, the change in renormalized chiral condensate eventually becomes negative with increasing $eB$ before rising again slightly. This reduction in the value of the chiral condensate with an increase in magnetic field strength $eB$ signifies the presence of an inverse magnetic catalysis effect. Notably, as the temperature increases, the slope of the initial rise diminishes, and the peak shifts toward lower values of $eB$. This shift in the peak position aligns with the similar reduction in $T_{pc}$ ~\footnote{{The pseudocritical temperature $T_{pc}$ can be defined in various ways due to the crossover nature of the transition. Throughout this work, $T_{pc}$ mainly refers to the inflection point of the renormalized chiral condensate with respect to temperature following the convention in~\cite{Bali:2011qj}.}} observed with increasing $eB$ \cite{Bali:2011qj}. Thus, the emergence of inverse magnetic catalysis coincides with the reduction of $T_{pc}$ as $eB$ increases~\footnote{It is important to note that when the pion mass in the vacuum exceeds approximately 550 MeV, the chiral condensates cease decreasing, a manifestation of the disappearance of inverse magnetic catalysis. However, even under these conditions, $T_{pc}$ continues to decrease with increasing $eB$~\cite{DElia:2018xwo, Endrodi:2019zrl}.}. 

At lower temperatures, $\Delta\Sigma_{s}$, like $\Delta\Sigma_{ud}$ and $\Delta\Sigma_{ds}$ increases with increasing $eB$ with a diminishing slope as the temperature increases. Unlike the lighter quark combination, $\Delta\Sigma_{s}$ does not reach a peak within the explored range of $eB$. At higher temperatures, the slope becomes negative, leading to an initial decrease in the value of $\Delta\Sigma_{s}$ as $eB$ increases. However, after reaching a minimum, $\Delta\Sigma_{s}$ begins to rise again with further increases in $eB$. 

\begin{figure}
\includegraphics[scale=0.63]{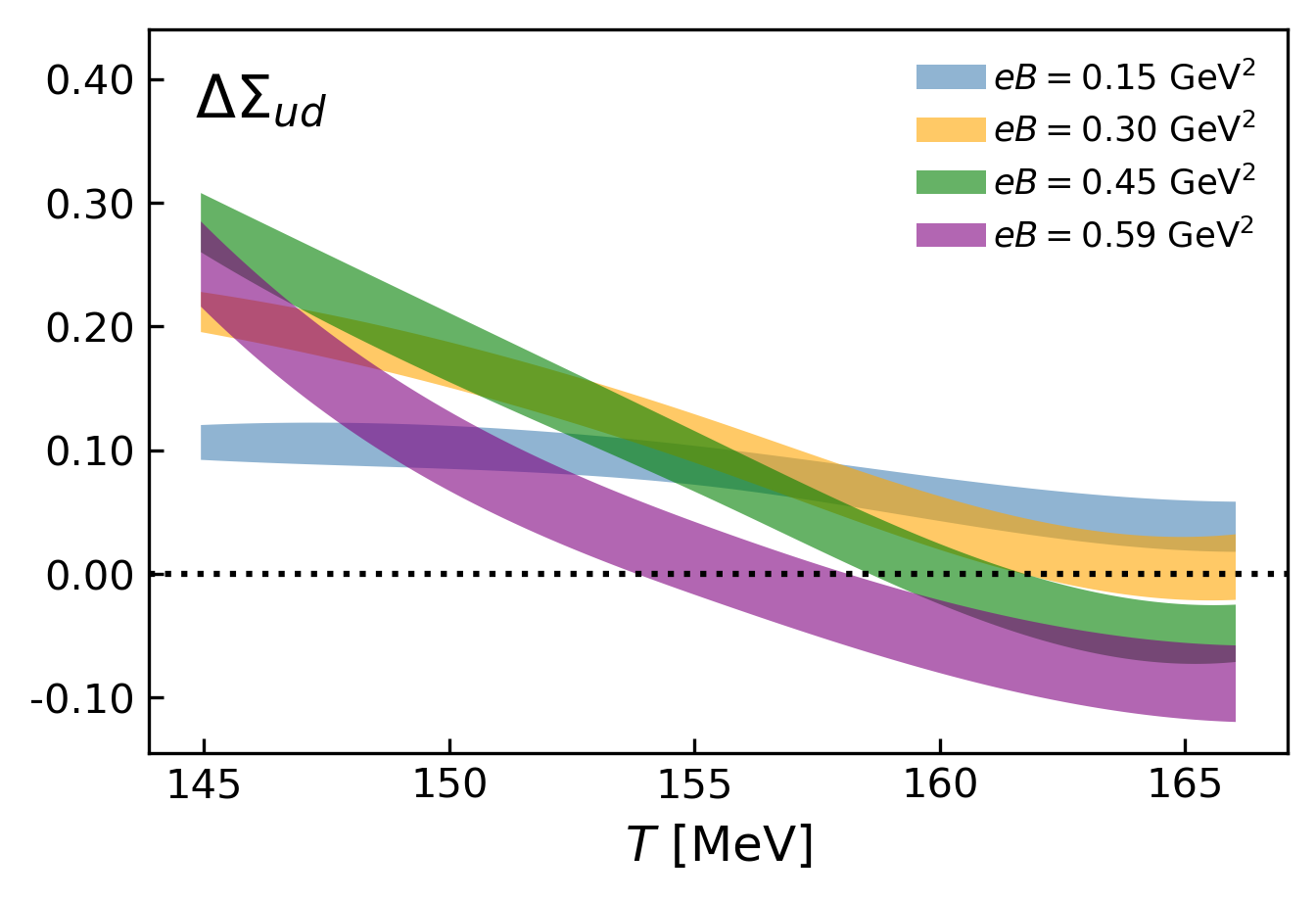}
\includegraphics[scale=0.63]{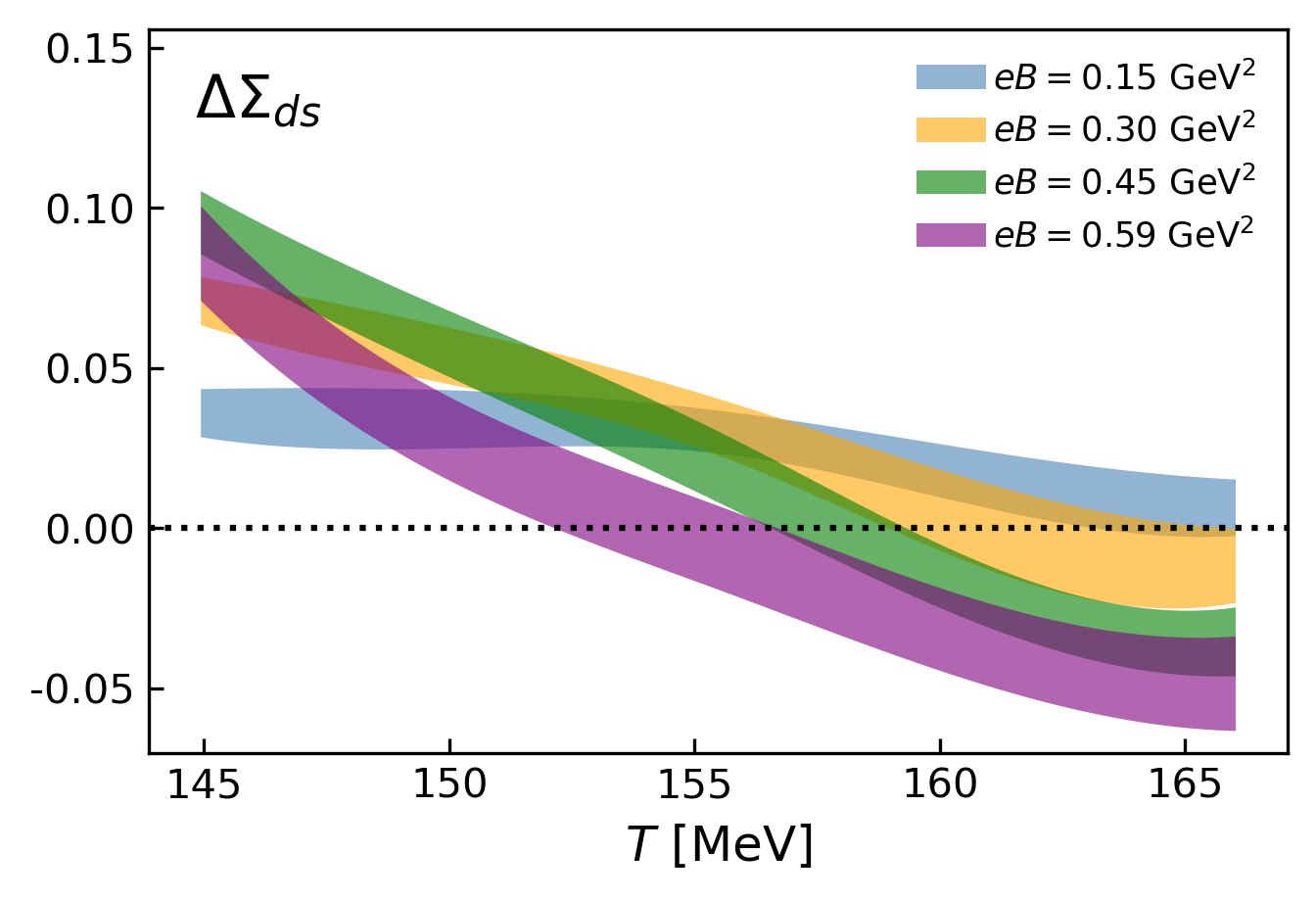}
\includegraphics[scale=0.63]{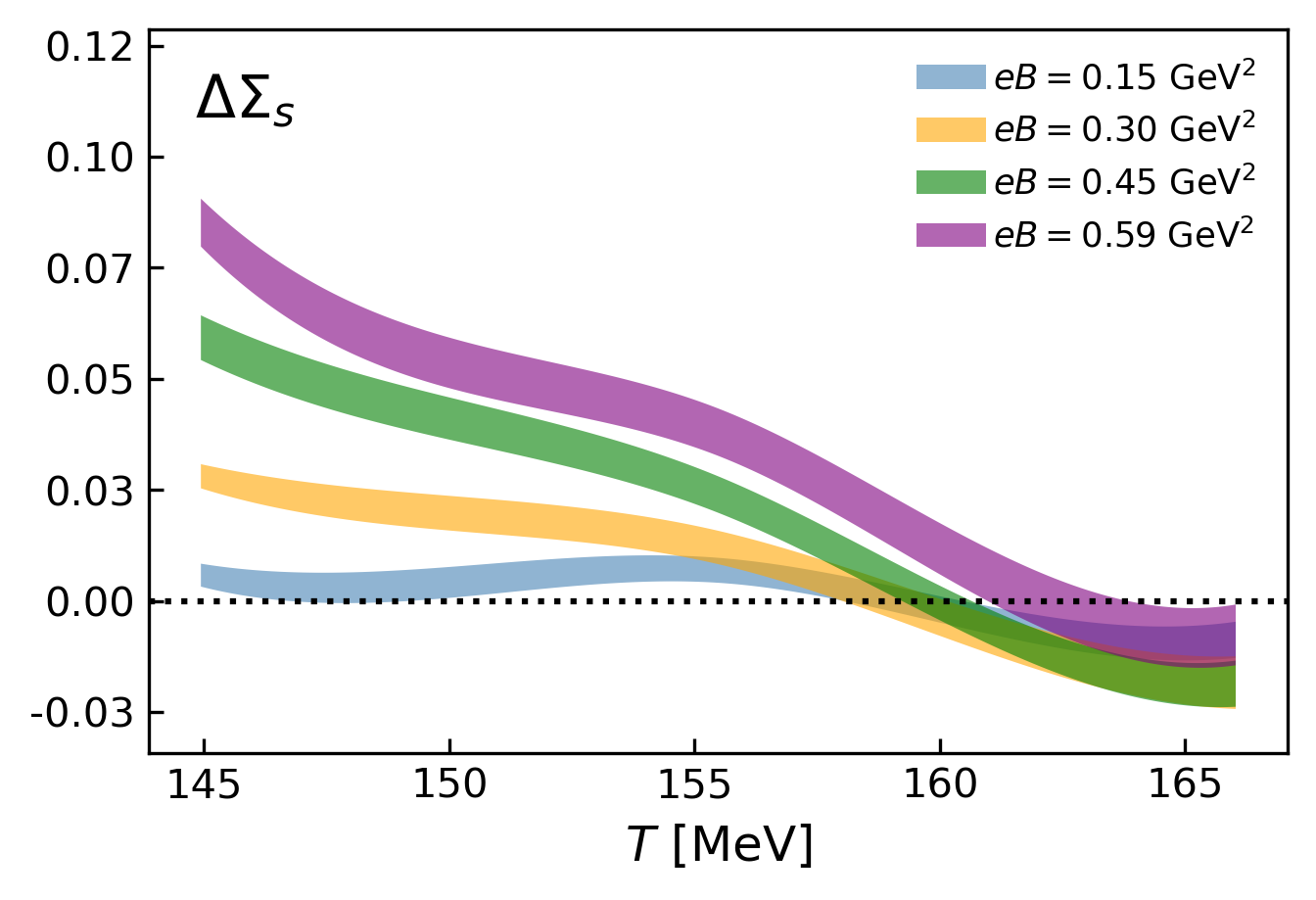}
\caption{ The continuum estimate of the change of the renormalized 
chiral condensates $\Delta\Sigma_{ud}$ (top), $\Delta\Sigma_{ds}$ (middle), $\Delta\Sigma_{s}$ (bottom) as a function of temperature
$T$ at fixed magnetic field strength $eB$. }
\label{pbp_vs_T}
\end{figure}

\autoref{pbp_vs_T} shows the temperature dependence of the changes in the renormalized chiral condensates: $\Delta\Sigma_{ud}$ (top), $\Delta\Sigma_{ds}$ (middle), and $\Delta\Sigma_{s}$ (bottom), at fixed values of the magnetic field strength, $eB$. Across all three quark combinations, the slope of the decline with increasing temperature becomes steeper for larger $eB$, leading to intersections between the curves for different magnetic field strengths. This behavior reflects the reduction in the pseudocritical temperature, $T_{pc}$, as noted earlier. However, at stronger magnetic fields, the initial values of the condensates at low temperatures do not necessarily increase. For example, in the case of the $ud$ and $ds$ condensates, the values at $eB = 0.59$ GeV$^2$ are lower than those at $eB = 0.45$ GeV$^2$. This trend is consistent with the decrease in these condensates observed at the two lowest temperatures in \autoref{pbp_vs_eb}. These findings suggest that at magnetic fields as strong as $eB \gtrsim 0.6$ GeV$^2$, the pseudocritical temperature, $T_{pc}$, appears to decrease to approximately 145 MeV~\cite{Bali:2011qj,DElia:2021yvk}.

The primary distinction between the quark combinations lies in the location where these intersections of the fixed $eB$ curves occur. For $\Delta\Sigma_{ud}$ and $\Delta\Sigma_{ds}$, the curves for the nonzero $eB$ curves intersect each other before crossing the $eB=0$ GeV$^2$ line. This intersection signifies the shift of dominance from the magnetic catalysis effect to the inverse magnetic catalysis effect. For $\Delta\Sigma_{s}$, the nonzero $eB$ curves first cross the $eB=0$ GeV$^2$ line, becoming negative before subsequently intersecting the other nonzero $eB$ curves as the temperature increases. We also show the $T$ and $eB$ dependence of the renormalized chiral condensates for $u$ and $d$ quarks separately in Appendix \ref{udquark}.

\begin{figure}
    \centering
        \includegraphics[width=0.95\linewidth]{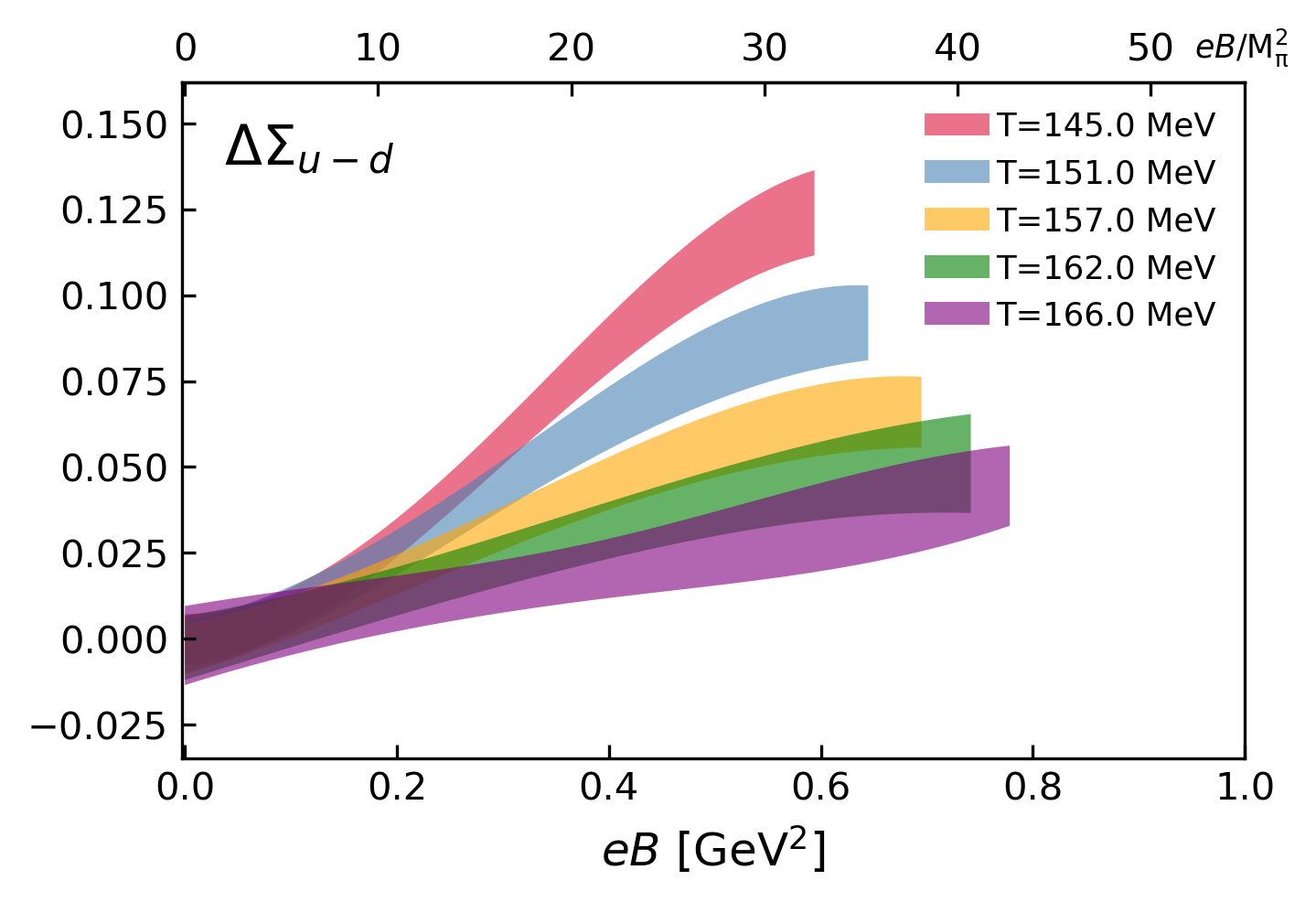}
    \includegraphics[width=0.9\linewidth]{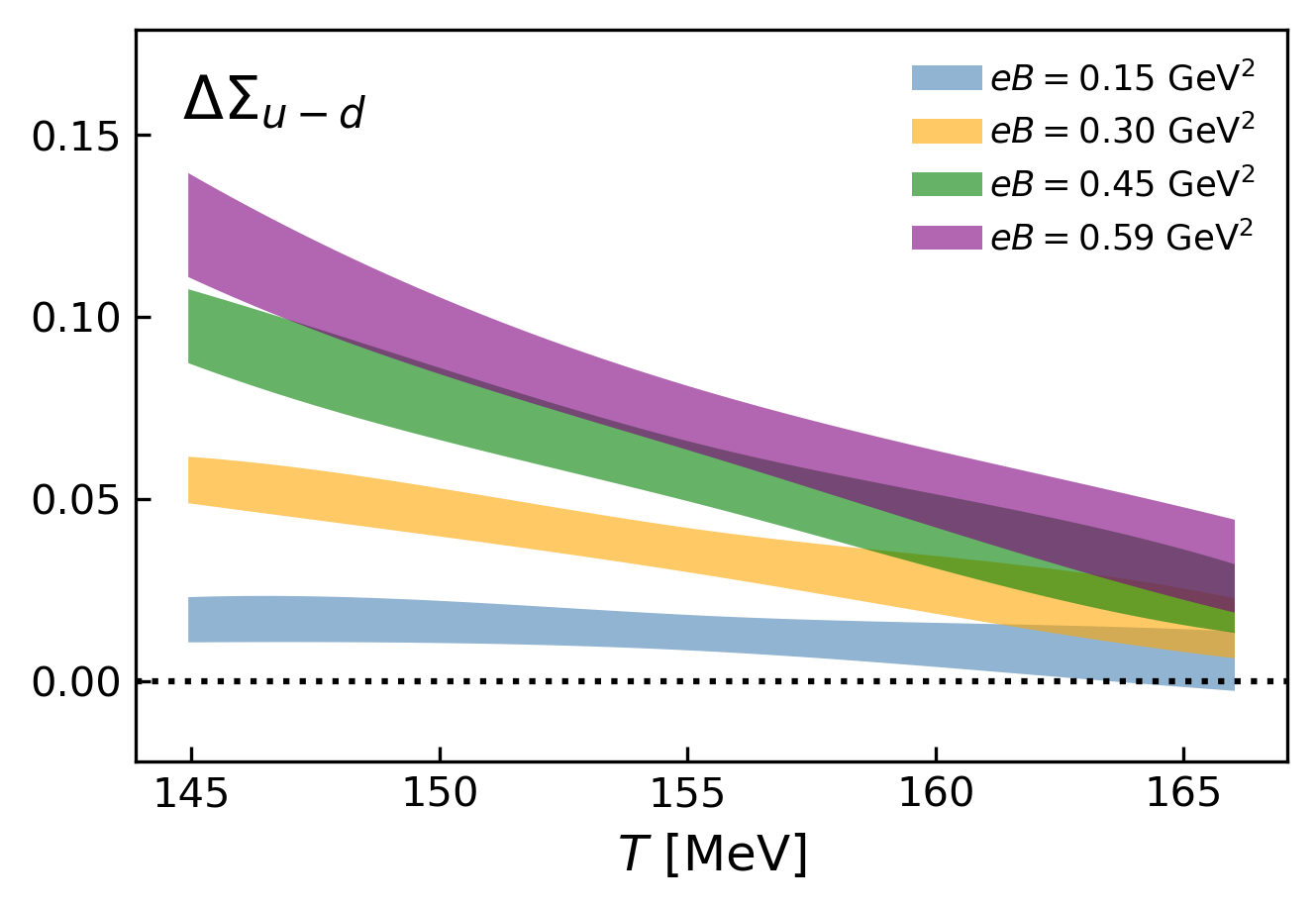}
    \caption{The continuum estimated difference of $u$ and $d$ quark chiral condensate $\Delta\Sigma_{u-d}$ as a function of (top) magnetic field and (bottom) temperature. }
    \label{u-d}
\end{figure}

\autoref{u-d} shows the difference in renormalized chiral condensates between the up and down quarks
\begin{equation}
    \Delta\Sigma_{u-d}(B,T)=\frac{m_u+m_d}{2M_\pi^2f_{\pi}^2}(\pbp_u(B,T)-\pbp_d(B,T))\,.
\end{equation}
In the top panel, $\Delta\Sigma_{u-d}$ is plotted as a function of the magnetic field strength at a fixed temperature. Across all temperatures, the difference between the up and down quark condensates grows with increasing $eB$. This trend suggests that the magnetic field enhances the asymmetry between the up and down quark condensates. As the temperature increases, the overall magnitude of this asymmetry decreases, reflecting a weakening of the effect of the magnetic field on the condensate difference at higher temperatures.

The bottom panel shows the variation of $\Delta\Sigma_{u-d}$ as a function of temperature at fixed values of the magnetic field. For nonzero $eB$, $\Delta\Sigma_{u-d}$  decreases monotonically with increasing temperature. The slope of this decline becomes steeper as $eB$ increases, indicating that the magnetic field enhances the chiral condensate difference more strongly at lower temperatures.

\begin{figure}
\includegraphics[scale=0.63]{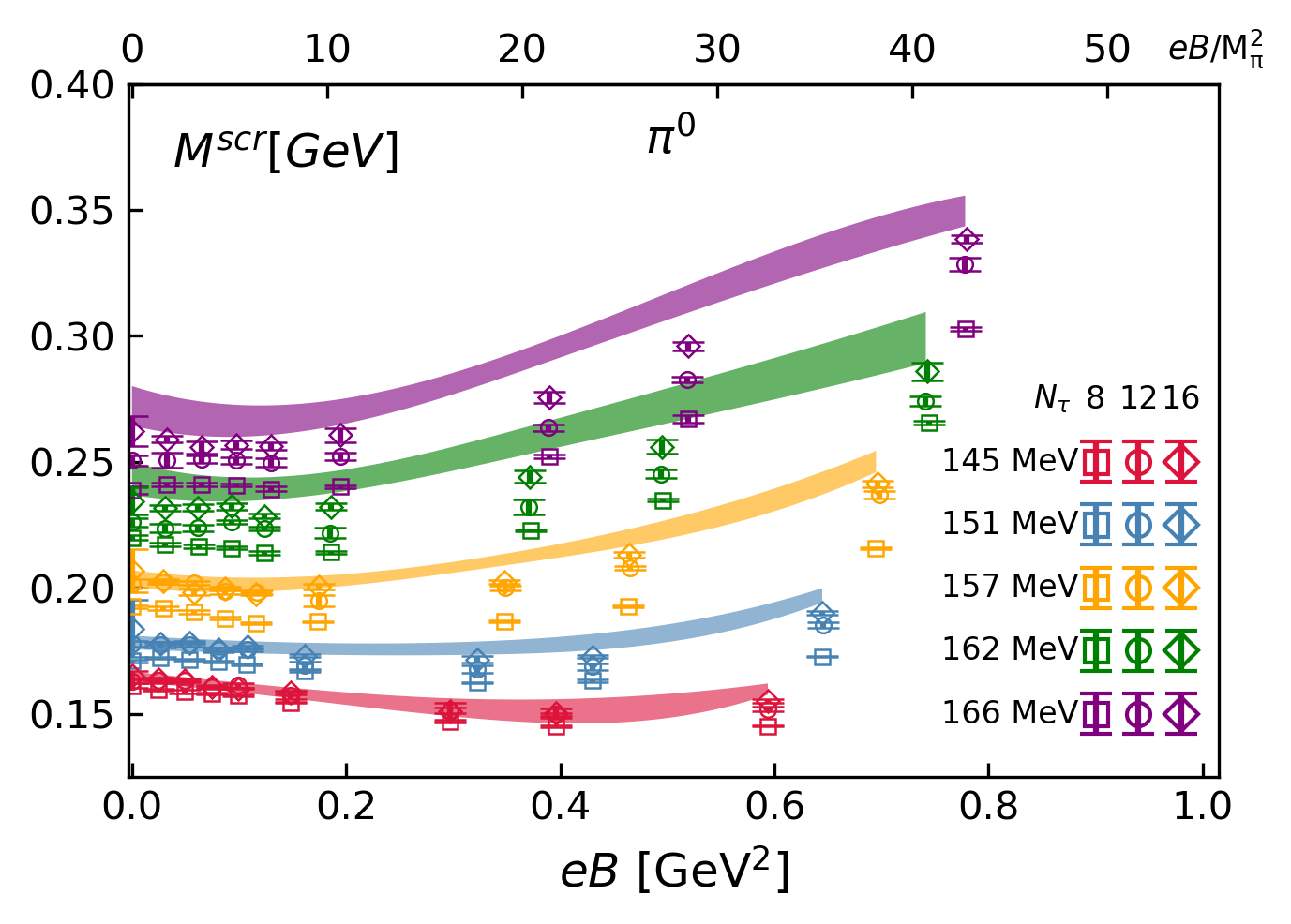}
\includegraphics[scale=0.63]{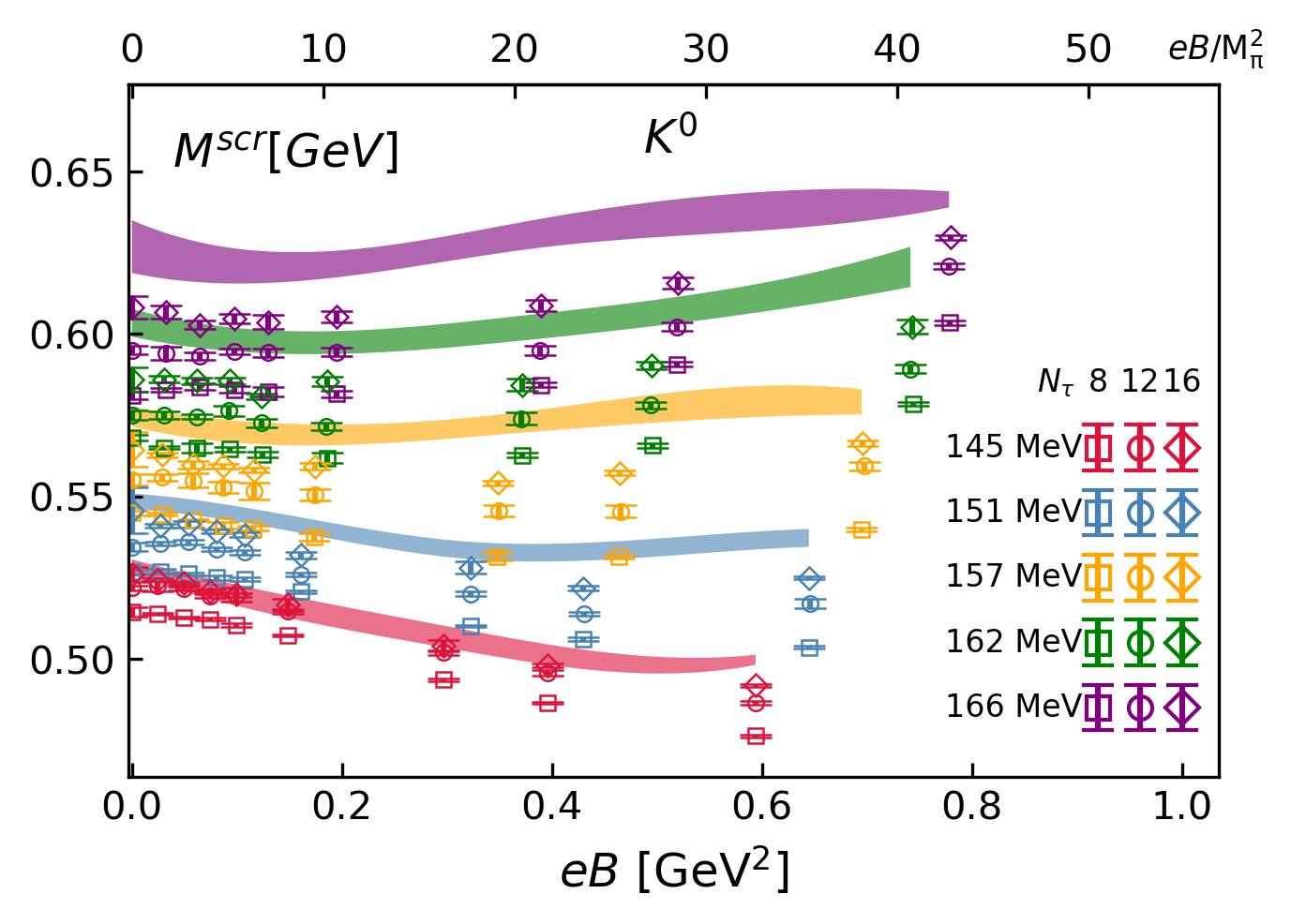}
\includegraphics[scale=0.63]{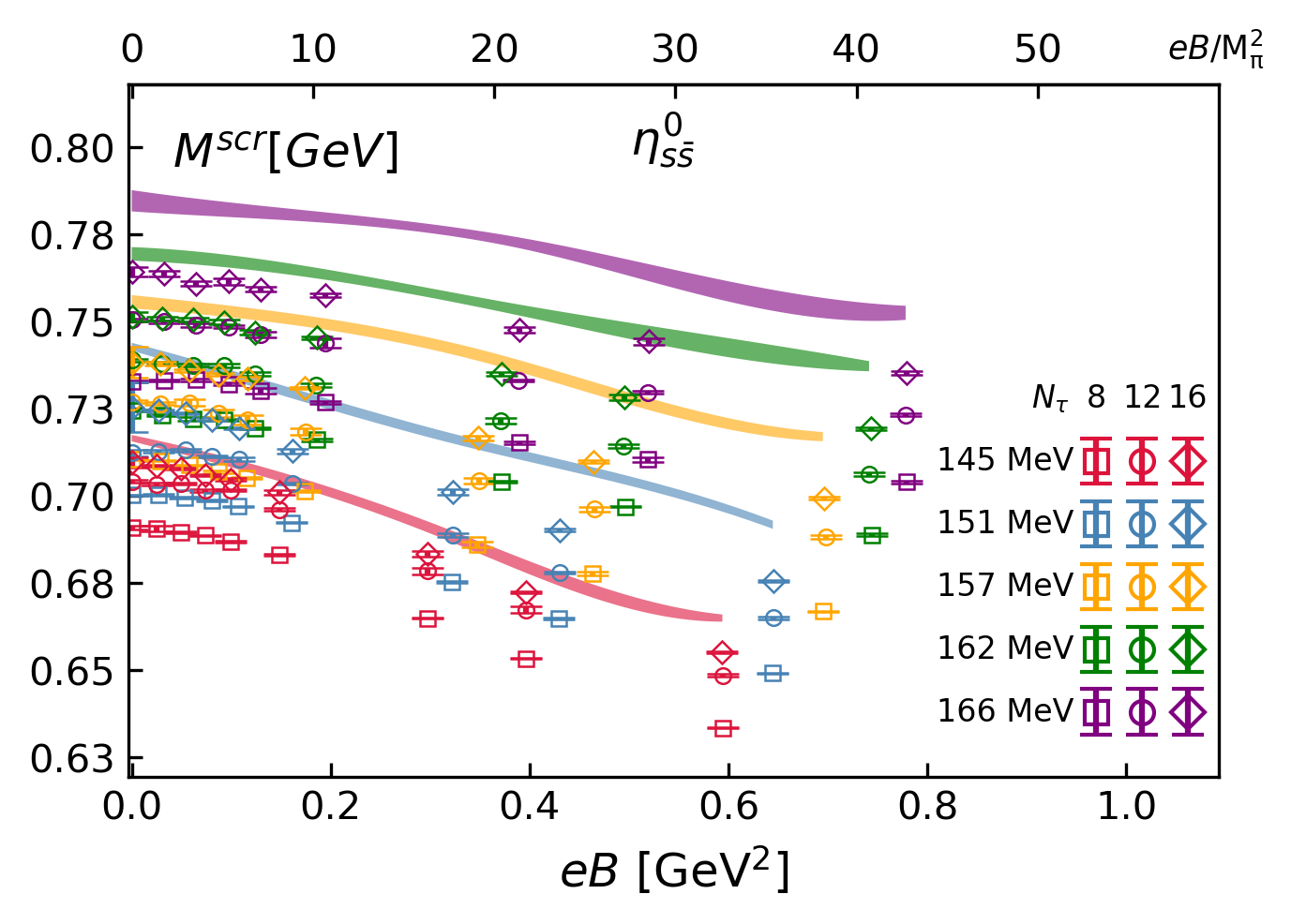}
\caption{The continuum estimate of the screening mass for neutral pseudoscalar mesons, namely $\pi^0$ (top), $K^0$ (middle), $\eta^0_{s\bar{s}}$ (bottom) as a function of magnetic field strength $eB$ at a fixed temperature
$T$. The shaded bands represent the continuum estimated results, while the data points correspond to lattice data at temperatures rounded to the nearest integer (refer to Appendix \ref{statistics} for exact temperature values). The upper $x$-axis is rescaled by the pion mass square in the vacuum at $eB = 0$ to make it dimensionless.}
\label{mscrn_vs_eb}
\end{figure}

\begin{figure}
\includegraphics[scale=0.63]{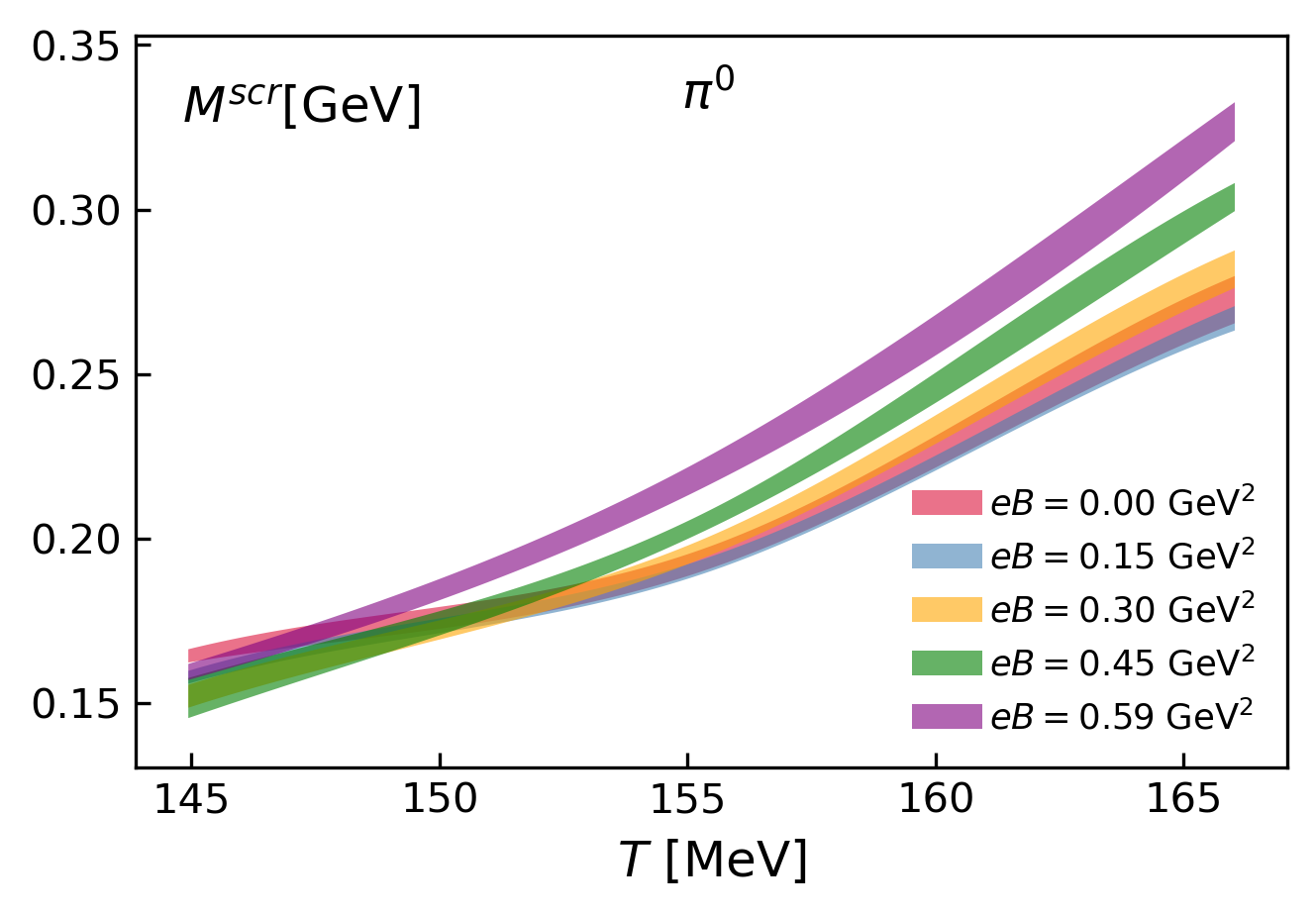}
\includegraphics[scale=0.63]{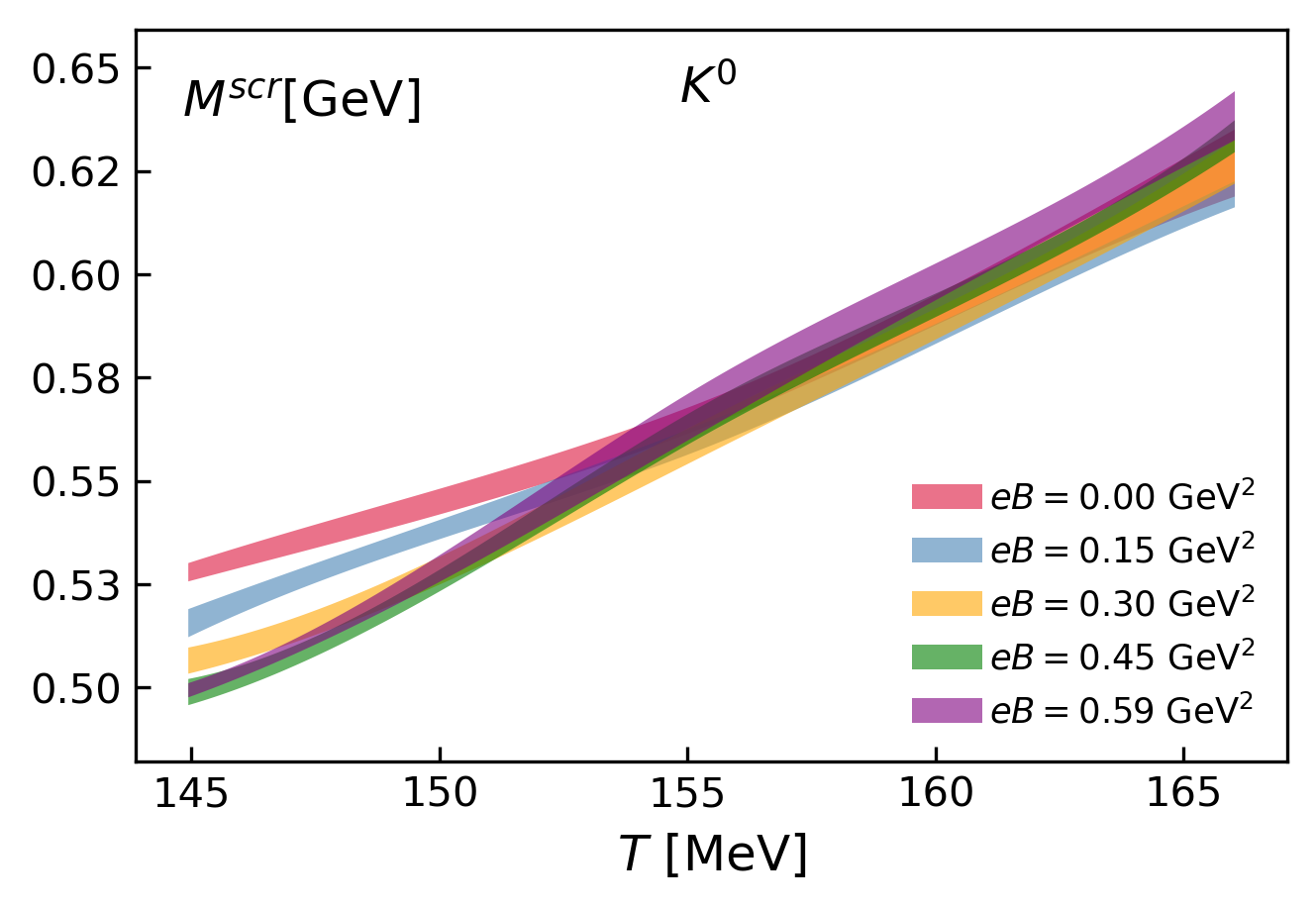}
\includegraphics[scale=0.63]{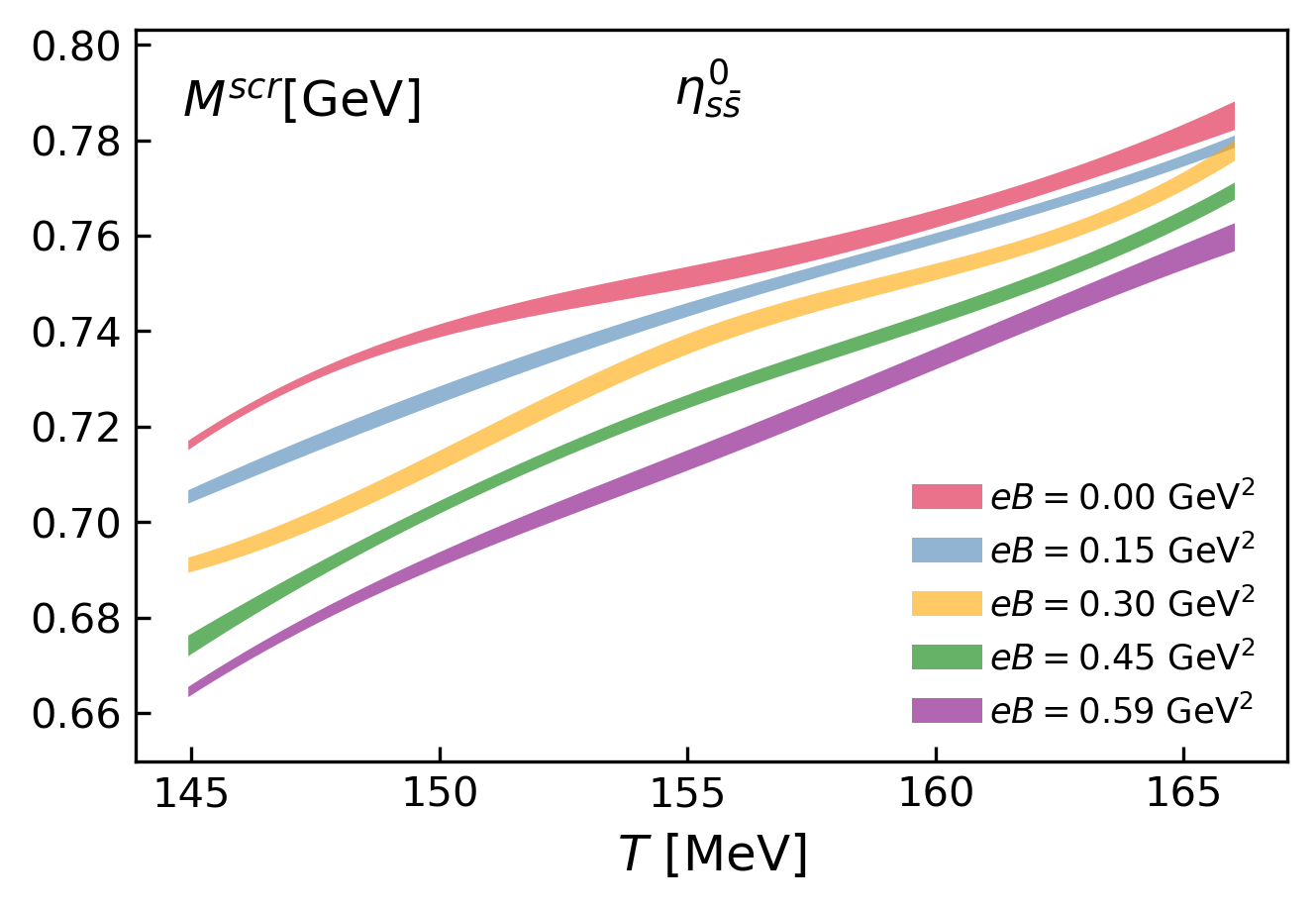}
\caption{ The continuum estimate of the screening mass for neutral pseudoscalar mesons, namely $\pi^0$ (top), $K^0$ (middle), $\eta^0_{s\bar{s}}$ (bottom) as a function of temperature
$T$ at fixed magnetic field strength $eB$. The band depicts the continuum estimate.}
\label{mscrn_vs_T}
\end{figure}

In \autoref{mscrn_vs_eb}, the continuum estimate of the screening masses for neutral pseudoscalar mesons,  $\pi^0$ \footnote{We assume equal  $\bar{u}u$ and $\bar{d}d$  contribution for the pion correlator, $G_{\pi^0}=\frac{G_{\bar{u}u}+G_{\bar{d}d}}{2}$. However, in a magnetic field, the $SU(2)_V$ symmetry is broken by the quarks' charges leading to mixing \cite{Bali:2017ian} of the light flavors. The neutral pion becomes a field-dependent combination, with both contributions becoming equal as $B\rightarrow0$.}(top),  $K^0$ (middle), and $\eta^0_{s\bar{s}}$ (bottom) mesons, are plotted as functions of the magnetic field strength $eB$ at different fixed temperatures $T$. Each band corresponds to continuum estimate at a distinct temperature value ranging from 145 MeV to 166 MeV. The dependence of screening masses on temperature and $eB$ is not necessarily identical to that of the corresponding chiral condensates, despite their connection through the relevant correlation functions (cf. Ward-Takahashi identities, \autoref{eq:Ward-id_pi}). This is because the screening mass reflects the asymptotic behavior of the meson correlation function at large spatial separations, thereby probing long-distance physics. In contrast, the chiral condensate, linked to the susceptibility, represents an integrated response over all distances, with a significant contribution from short-distance correlations due to the exponential decay of the correlator. Thus, while both observables are sensitive to chiral symmetry breaking, they capture distinct physical aspects of the thermomagnetic medium.

Previous studies at zero temperature \cite{Bali:2017ian,Ding:2020hxw,Luschevskaya:2014lga,Luschevskaya:2015cko} have shown that the neutral meson screening mass decreases monotonically with increasing magnetic field $eB$, eventually reaching a plateau. At low temperatures of our analysis, the screening mass of the neutral pion $\pi^0$ exhibits non-monotonic behavior, initially decreasing to a minimum before rising as $eB$ increases. The minima shift towards smaller $eB$ values as the temperature increases. This effect was explained by effective model \cite{Sheng:2021evj} and our previous paper for heavier quark mass \cite{Ding:2022tqn} where it was observed that the sea quarks induce an inverse magnetic catalysis causing the screening mass to increase with the magnetic field. Similar behavior is also observed for the neutral kaon $K^0$, though with more pronounced minima, especially for lower temperatures. Also, the slopes at higher values of $eB$ are gentler for $K^0$ when compared with $\pi^0$. The behavior of the fictitious eta meson $\eta^0_{s\bar{s}}$ is different from the other two mesons and shows a steady decrease in magnitude with increasing magnetic field $eB$ with the slope more pronounced for lower temperatures. It can also be seen that the heavier the meson mass is the less is affected by the magnetic field. This is consistent with the finding from simulations with larger-than-physical pion mass~\cite{Ding:2022tqn}.

In \autoref{mscrn_vs_T}, we show the continuum estimate of screening masses for neutral pseudoscalar mesons for $\pi^0$ (top), $K^0$ (middle), and the fictitious $\eta^0_{s\bar{s}}$ meson (bottom) — as a function of temperature $T$ for various fixed strengths of the magnetic field $eB$. For all three mesonic channels, the screening masses increase monotonically with temperature across all values of $eB$. This is a result of thermal effects predominantly enhancing the screening masses, indicating shifts in mesonic properties as the system approaches the pseudo-critical temperature $T_{pc}$ associated with chiral symmetry restoration.

For the neutral pion $\pi^0$, the screening mass curves exhibit a steeper slope at larger $eB$ compared to smaller $eB$, and they appear to diverge at higher temperatures. At lower temperatures, the constant $eB$ curves intersect one another, with the crossing point of the $eB = 0$ curve occurring at a lower temperature for larger $eB$ values. This behavior suggests a reduction in the pseudo-critical temperature $T_{\text{pc}}$ in stronger magnetic fields. This crossing feature resembles the behavior observed in the chiral condensate shown in \autoref{pbp_vs_T}, where stronger magnetic fields shift $T_{\text{pc}}$ to lower values. A similar intersecting behavior is observed for the neutral kaon $K^0$, although the crossover occurs at relatively higher temperatures compared to $\pi^0$. In contrast, the fictitious $\eta^0_{s\bar{s}}$ meson does not exhibit such intersections in the covered temperature window; instead, the curves for different $eB$ values appear to intersect at sufficiently high temperatures, beyond the current temperature window.

\section{CONCLUSION}
\label{conclusion}

In this work, we conducted a comprehensive study on the effects of temperature and magnetic field on the chiral condensates and screening masses of neutral pseudoscalar mesons near the pseudocritical temperature, $T_{pc}$. Using lattice QCD simulations with physical quark masses, we performed continuum estimations with three lattice spacings to minimize discretization effects. Although chiral condensates and mesonic screening masses are related through the Ward-Takahashi identity, their dependencies on $T$ and $eB$ exhibit intricate, non-trivial behaviors due to the competing mechanisms of magnetic catalysis and inverse magnetic catalysis.

At $T = 145$, $151$, and $157$ MeV, the light ($ud$) and strange-light ($ds$) quark chiral condensates initially increase with $eB$, followed by a decrease at sufficiently strong magnetic fields. In contrast, the strange quark ($s$) chiral condensate increases monotonically with $eB$ at these temperatures. At the higher temperatures of $162$ and $166$ MeV, the light and strange-light condensates exhibit a more complex pattern: they initially increase, then decrease, and finally tend to rise again as $eB$ grows. For the strange quark condensate at these temperatures, the behavior reverses, with an initial decrease followed by an increase at stronger magnetic fields. Across all temperatures, chiral condensates decrease with increasing $T$, and this suppression is more pronounced at stronger magnetic fields and for lighter quarks.

For the screening masses of neutral pseudoscalar mesons, we observe non-monotonic behavior in $\pi^0$ and $K^0$, which are linked to the light and strange-light chiral condensates, respectively. In contrast, the screening mass of $\eta^0_{s\bar{s}}$ decreases monotonically with increasing $eB$. In terms of temperature dependence, screening masses consistently increase with $T$ across all magnetic field strengths, with steeper slopes observed at stronger magnetic fields. Notably, $\pi^0$ and $K^0$ exhibit intersecting behavior in this context. These temperature dependencies of the screening masses further support the observation that an external magnetic field reduces the pseudocritical temperature, $T_{pc}$. However, this reduction of $T_{pc}$ is less manifested in the behavior of mesons composed of heavier quarks, reflecting the characteristics of the chiral crossover transition within the explored temperature and magnetic field window. All data presented in the figures of the paper can be found in \cite{thakkar}.

\vspace{0.0cm}
\section{ACKNOWLEDGEMENTS}
This research was partially funded by the National Natural Science Foundation of China under Grants No. 12293060, No. 12293064; No. 12325508, along with support from the National Key Research and Development Program of China under Contract No. 2022YFA1604900. Computational resources for the numerical simulations were provided by the GPU cluster at the Nuclear Science Computing Center, Central China Normal University (NSC$^3$), and the Wuhan Supercomputing Center.

\bibliography{ref}
\bibliographystyle{JHEP.bst}

\appendix
\renewcommand{\appendixautorefname}{Appendix}
\section{Spline Interpolation and Continuum estimate}
\label{interpolation}
\def\sectionautorefname{Appendix}

\begin{figure}
\includegraphics[scale=0.57]{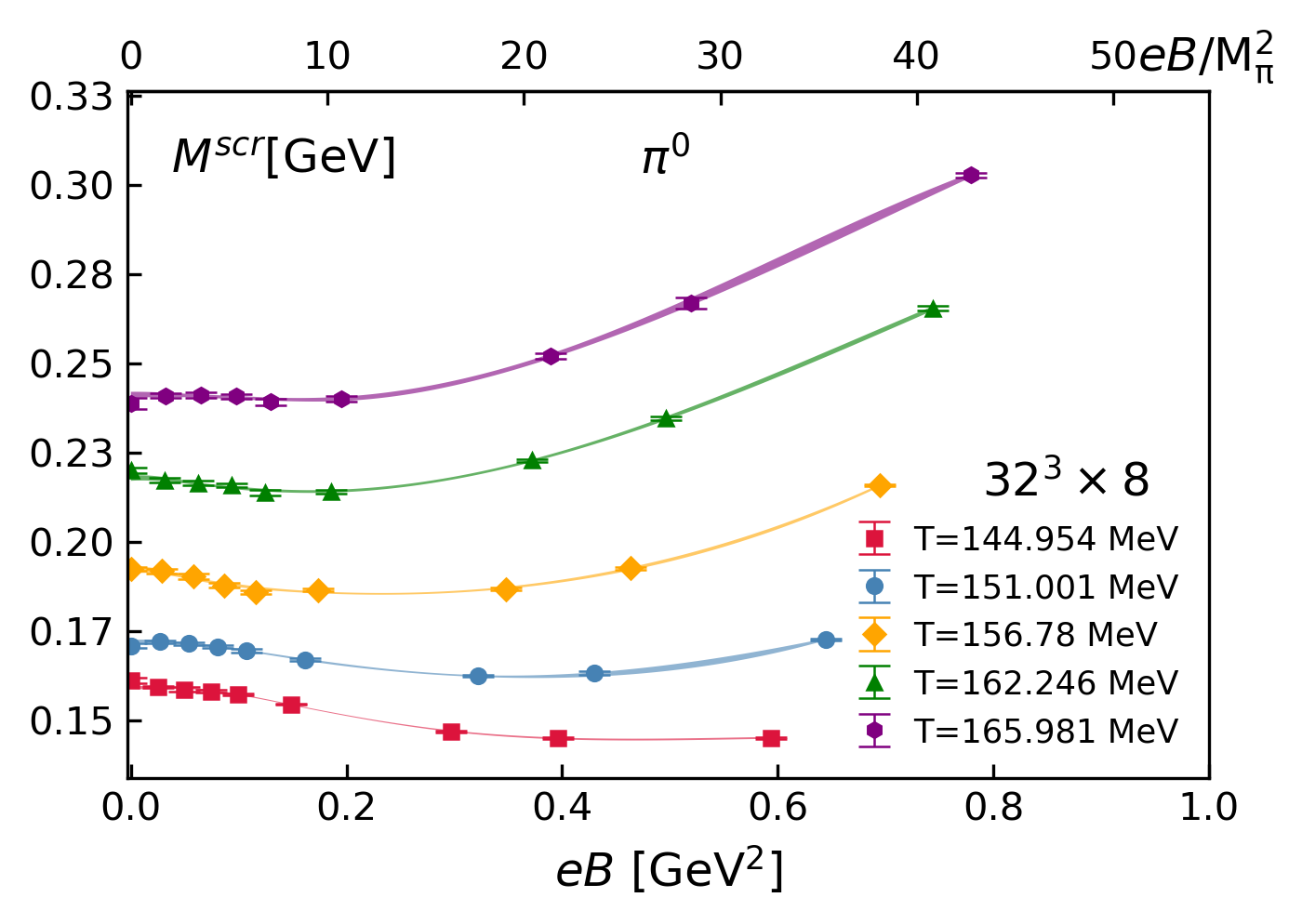}
\includegraphics[scale=0.57]{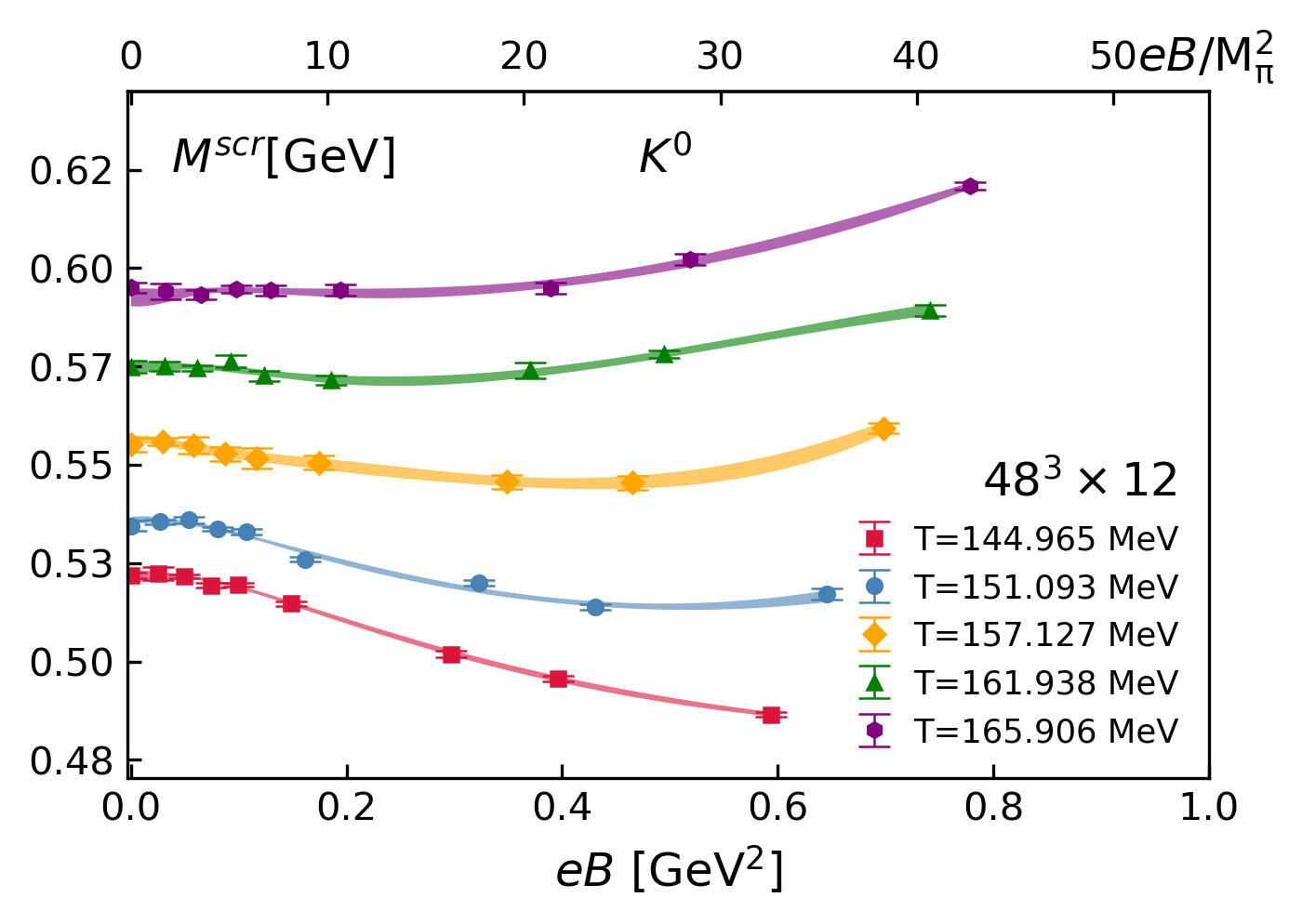}
\includegraphics[scale=0.57]{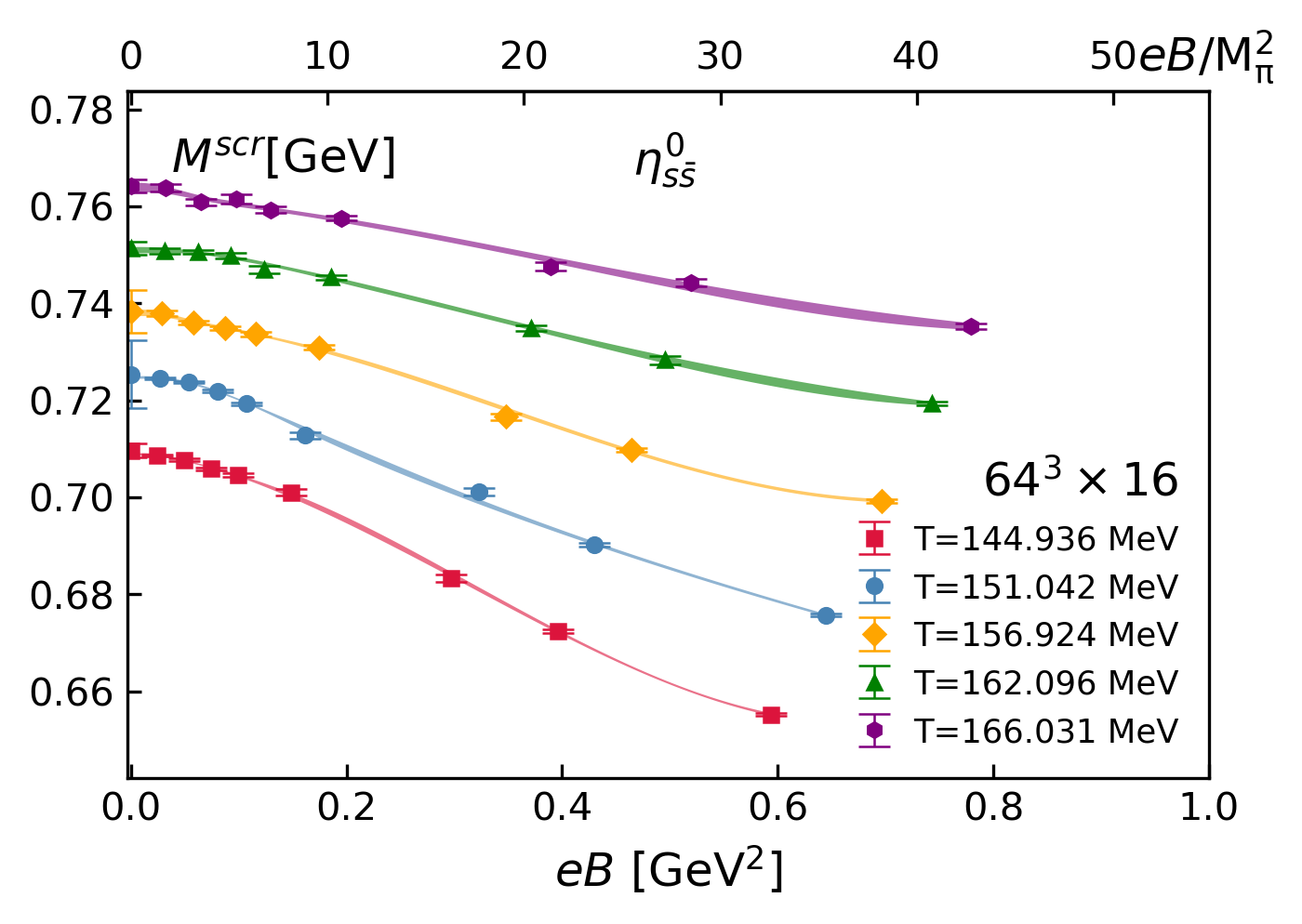}
\caption{ Samples of interpolation estimate of the screening mass as a function of the magnetic field strength
$eB$ at fixed temperatures are shown for three quark-lattice dimension combinations:  $32^3\times 8$ for $\pi^0$ (top), $48^3\times 12 $ for $K^0$ (middle), $64^3\times 16$ for $\eta^0_{s\bar{s}}$ (bottom). The data points are the lattice data and the bands depict the interpolated values in the $T - eB$ plane. The upper $x$-axis is rescaled by the pion mass square in the vacuum at $eB = 0$ to make it dimensionless.}
\label{fig:interpolation}
\end{figure}

\begin{figure}[t!]
{
\includegraphics[width=0.42\textwidth]{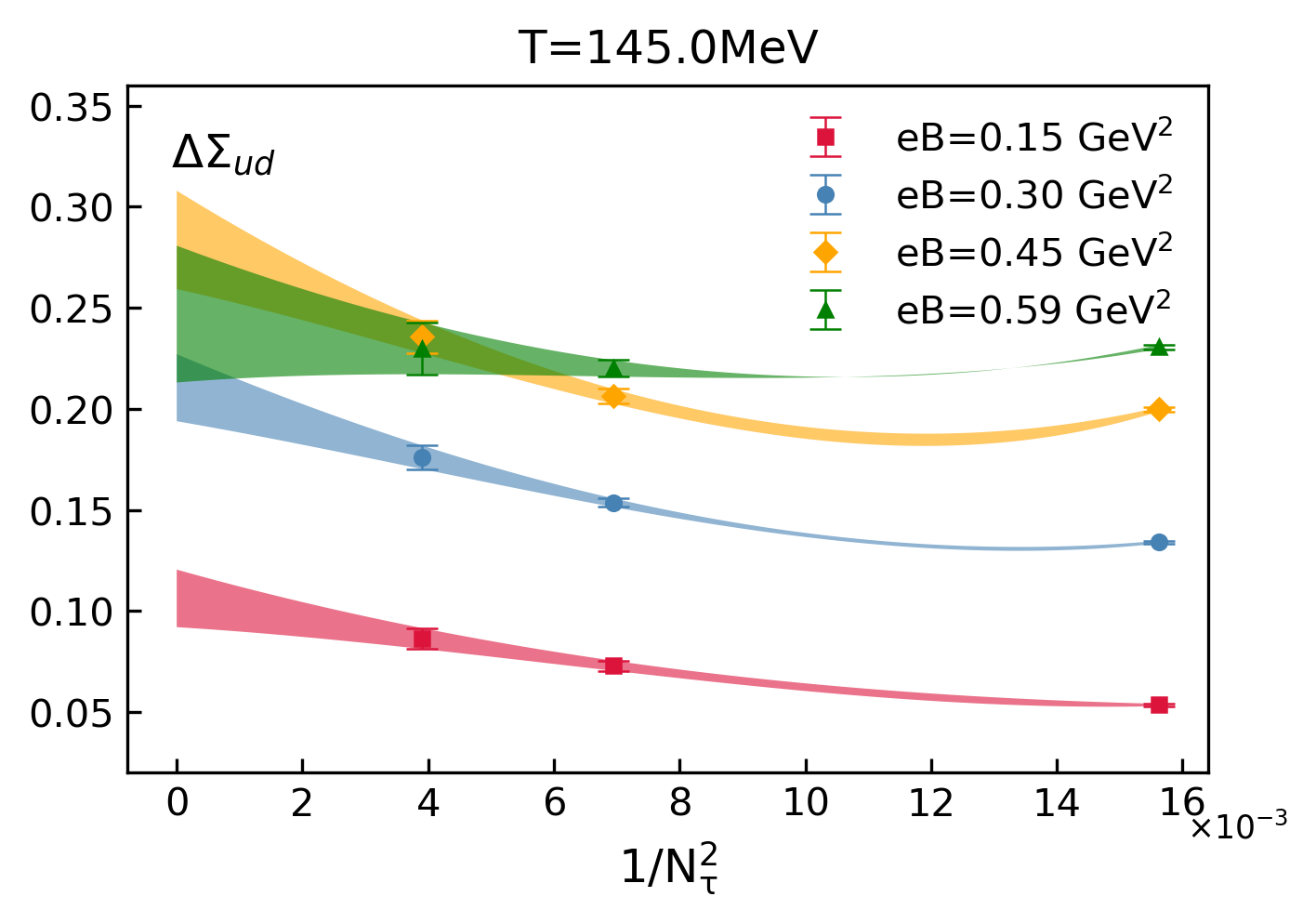}
\includegraphics[width=0.42\textwidth]{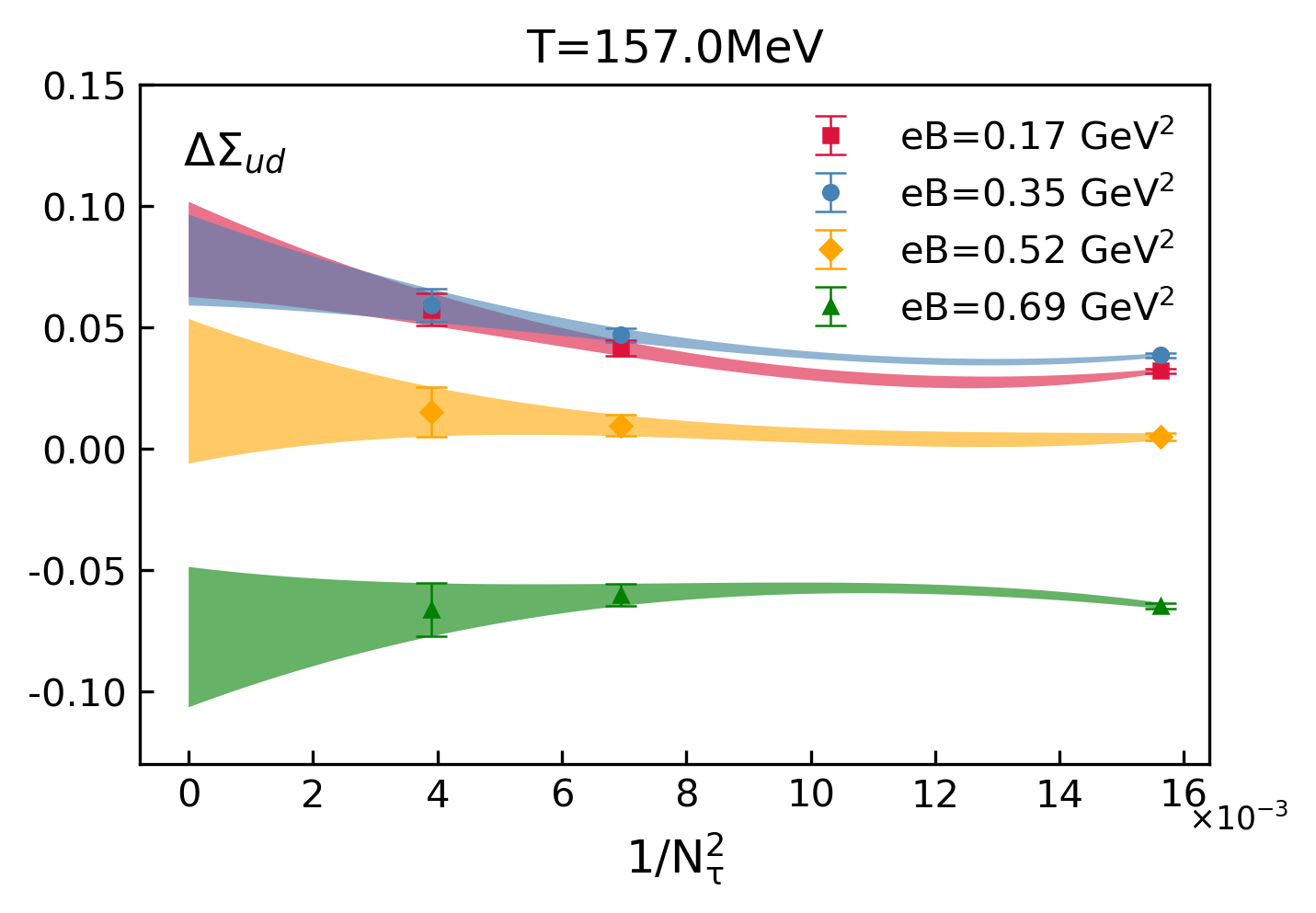}
\includegraphics[width=0.42\textwidth]{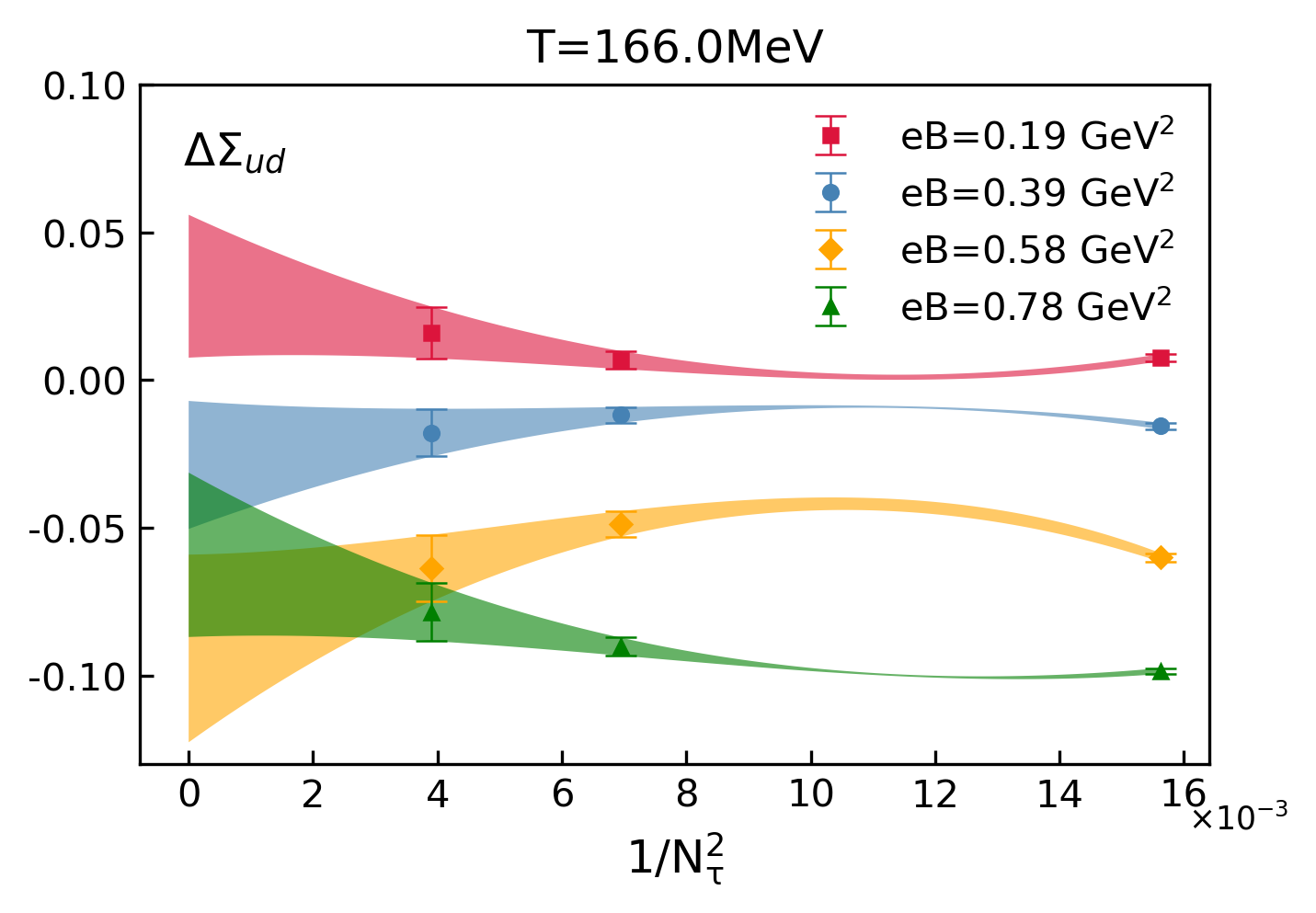}
\caption{ Samples of continuum estimate of the difference of light chiral condensate $\Sigma_{ud}$ at fixed temperatures $T$ and magnetic field strength $eB$ are shown for three temperatures $T$=145.0 MeV (top),  $T$=157.0 MeV (middle), and at $T$=166.0 MeV (bottom). The data points are obtained through interpolation of the lattice data and the bands depict the continuum estimate error.}
\label{sigma_estimate}
}
\end{figure}

\begin{figure}[!t]
{
\includegraphics[width=0.42\textwidth]{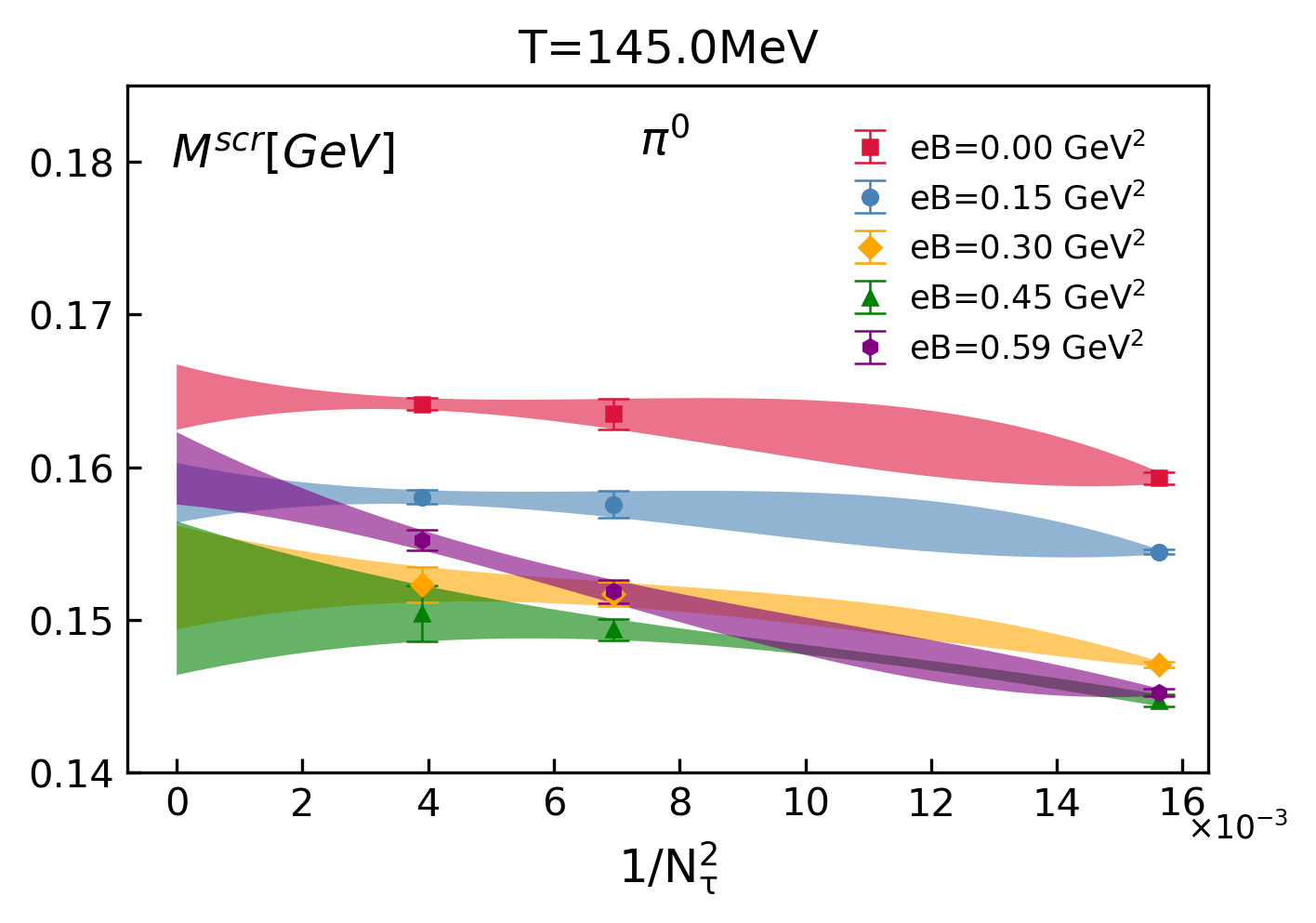}
\includegraphics[width=0.42\textwidth]{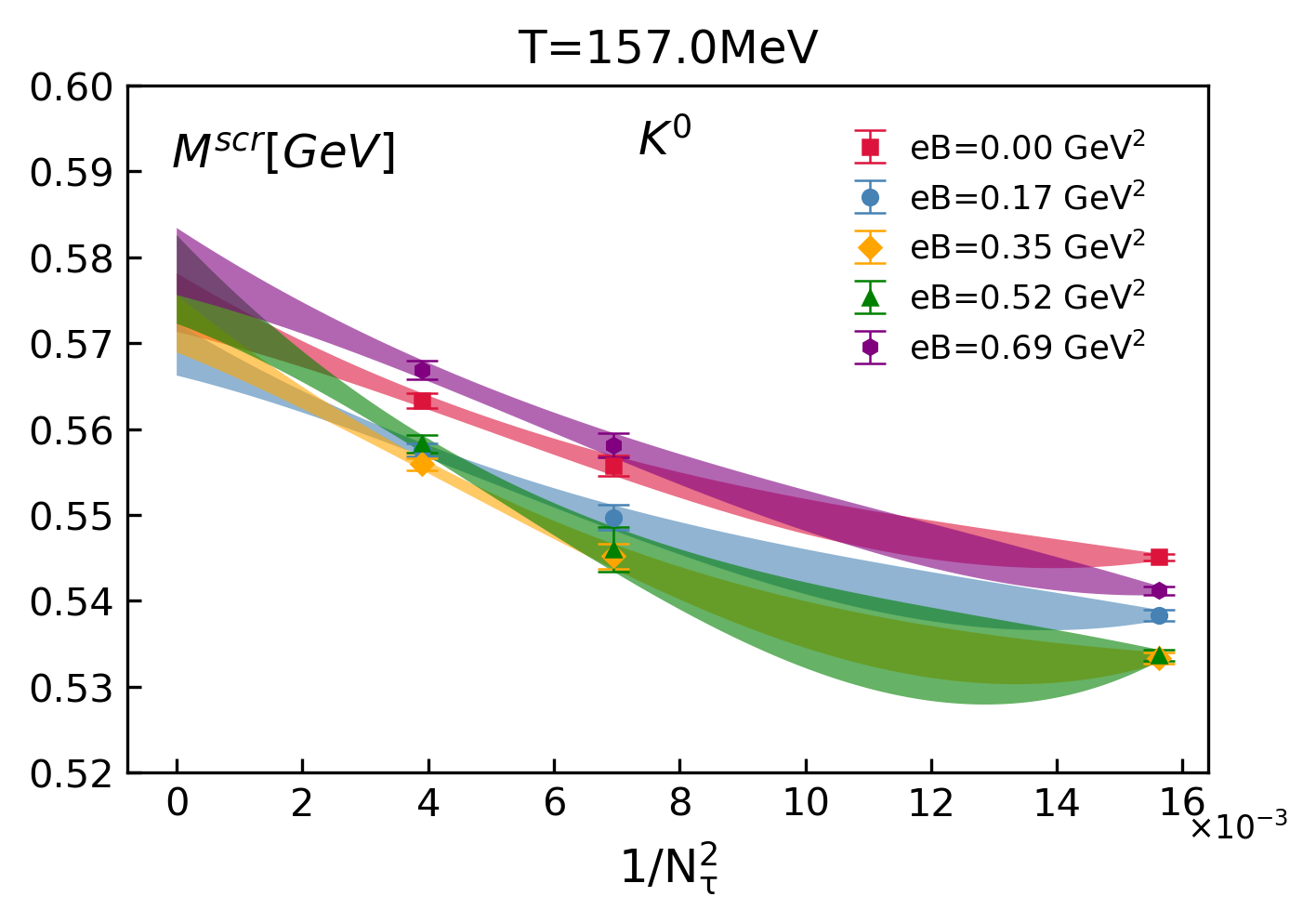}
\includegraphics[width=0.42\textwidth]{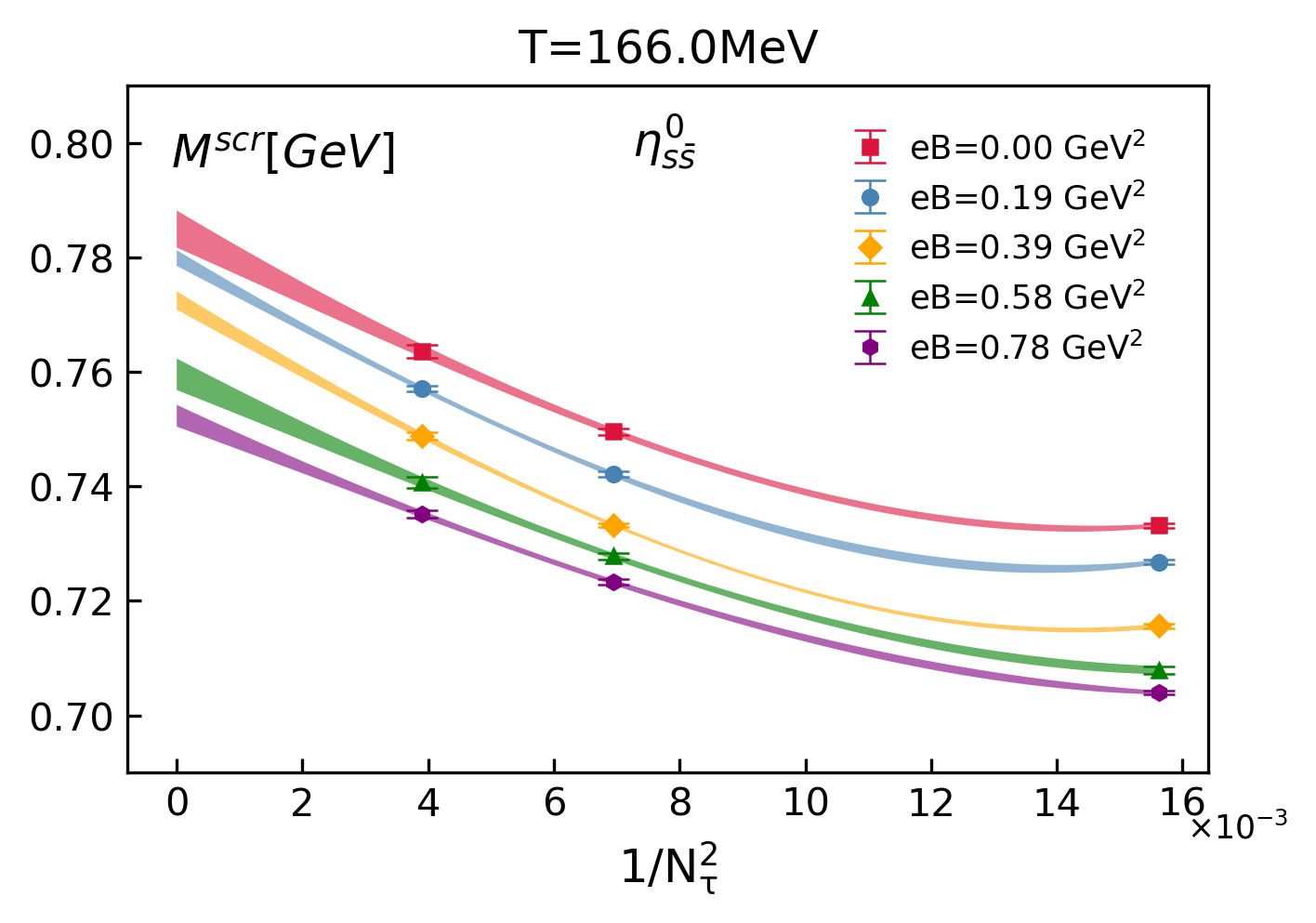}
\caption{ Samples of continuum estimate of the screening mass at fixed temperatures $T$ and magnetic field strength $eB$ are shown for three temperatures and quark combination: $\pi^0$ at $T$=145.0 MeV (top), $K^0$ at $T$=157.0 MeV (middle), and $\eta^0_{s\bar{s}}$ at $T$=166.0 MeV (bottom). The data points are obtained through interpolation of the lattice data and the bands depict the continuum estimate error.}
\label{estimate}
}
\end{figure}
\begin{figure}
\includegraphics[scale=0.67]{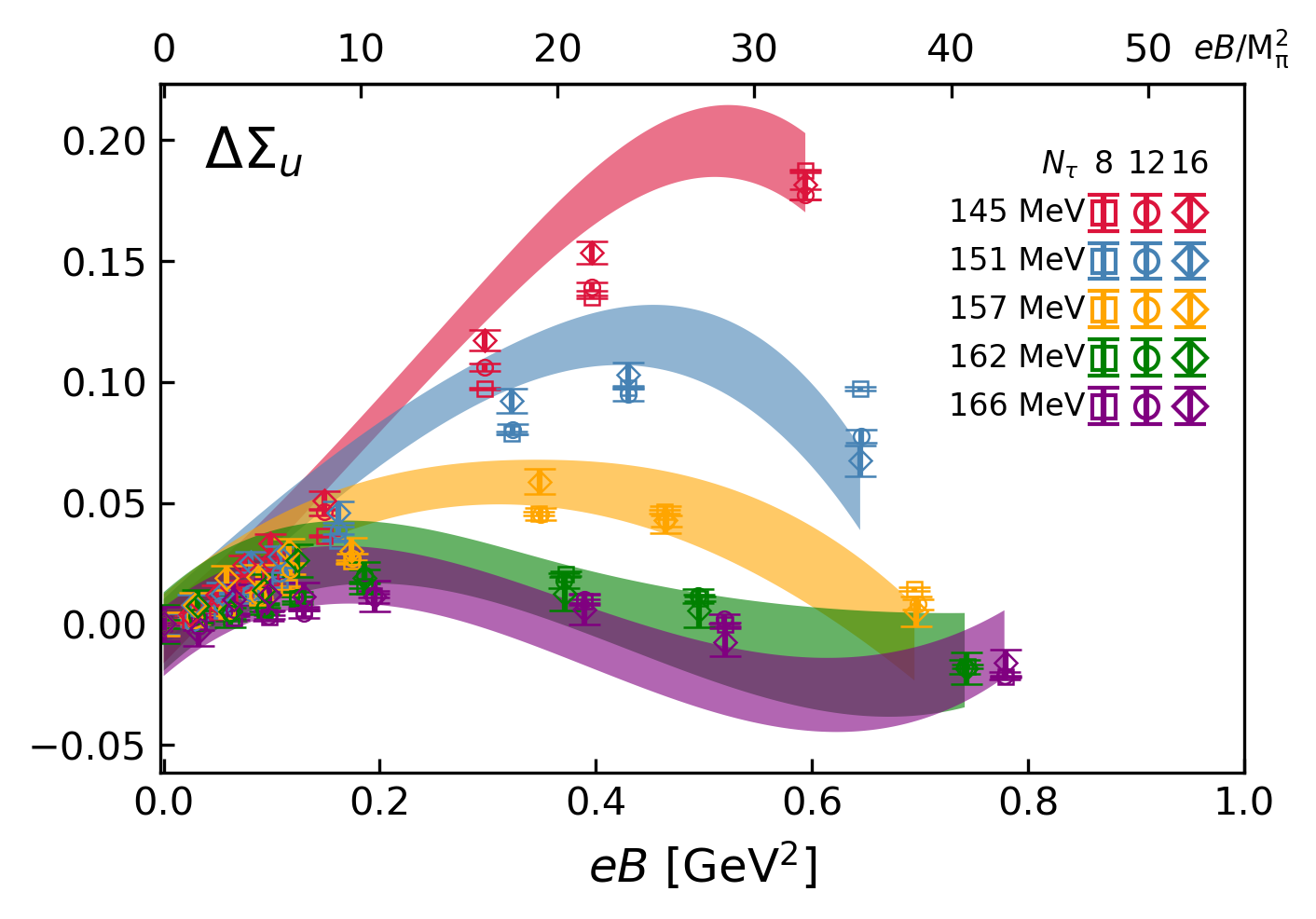}
\includegraphics[scale=0.67]{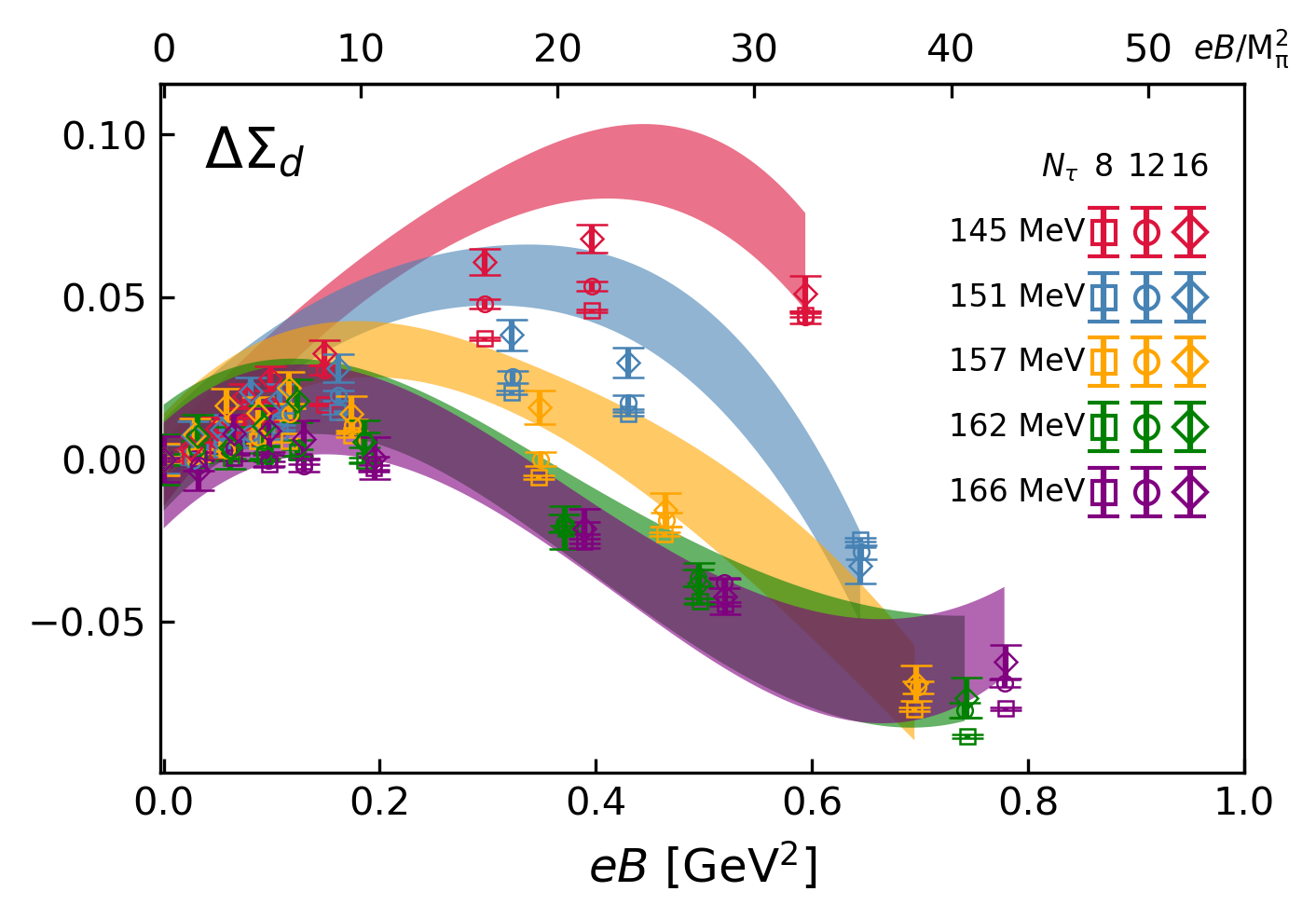}
\caption{The continuum estimate of the change of the renormalized 
chiral condensates $\Delta\Sigma_{u}$ (top) and $\Delta\Sigma_{d}$ (bottom) as a function of the magnetic field strength
$eB$ at fixed temperatures. The shaded bands represent the continuum estimated results, while the data points correspond to lattice data at temperatures rounded to the nearest integer(refer to Appendix \ref{statistics} for exact temperature values). The upper $x$-axis is rescaled by the pion mass square in the vacuum at $eB = 0$ to make it dimensionless.
}
\label{pbpq_vs_eb}
\end{figure}

\begin{figure}
\includegraphics[scale=0.67]{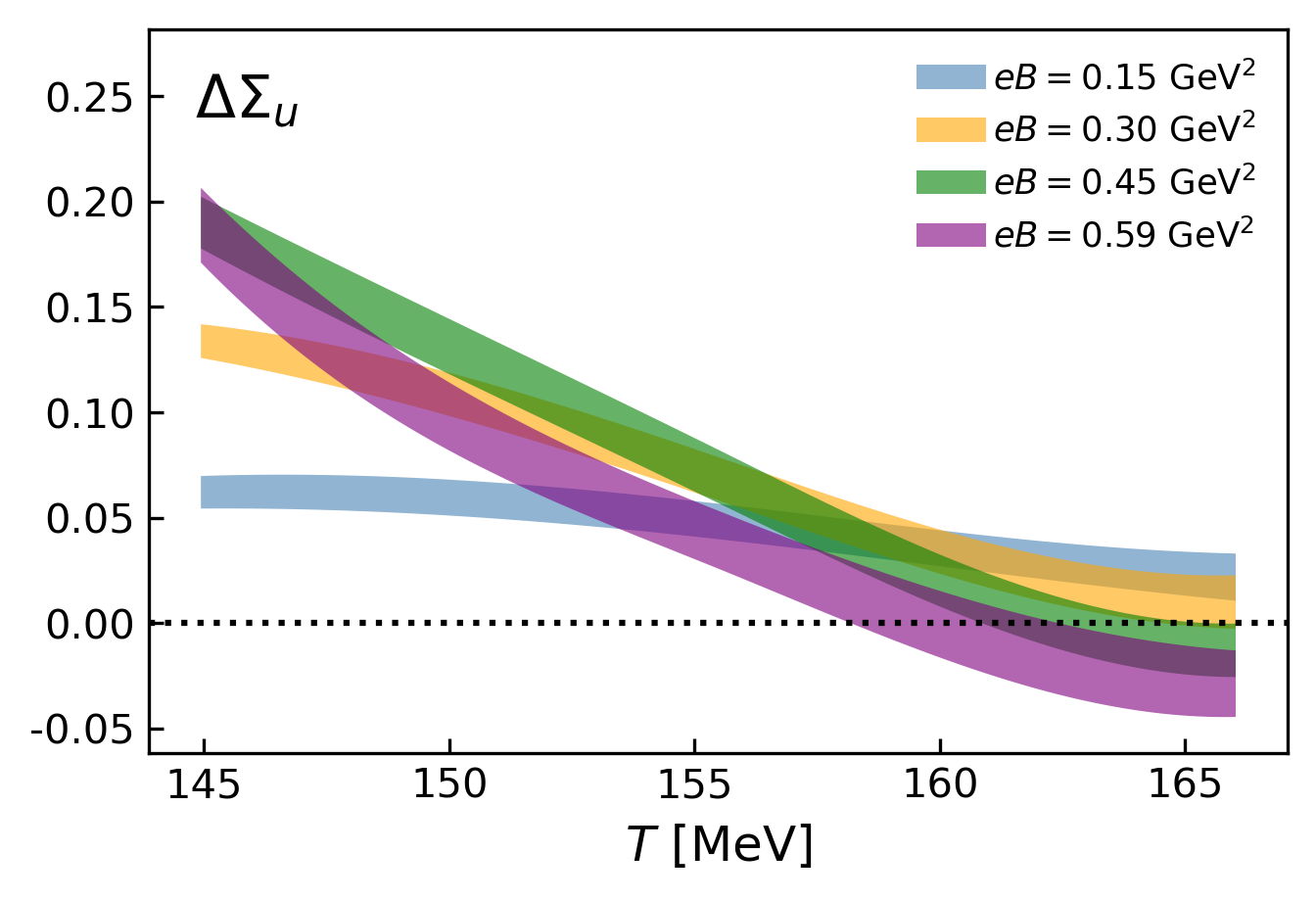}
\includegraphics[scale=0.67]{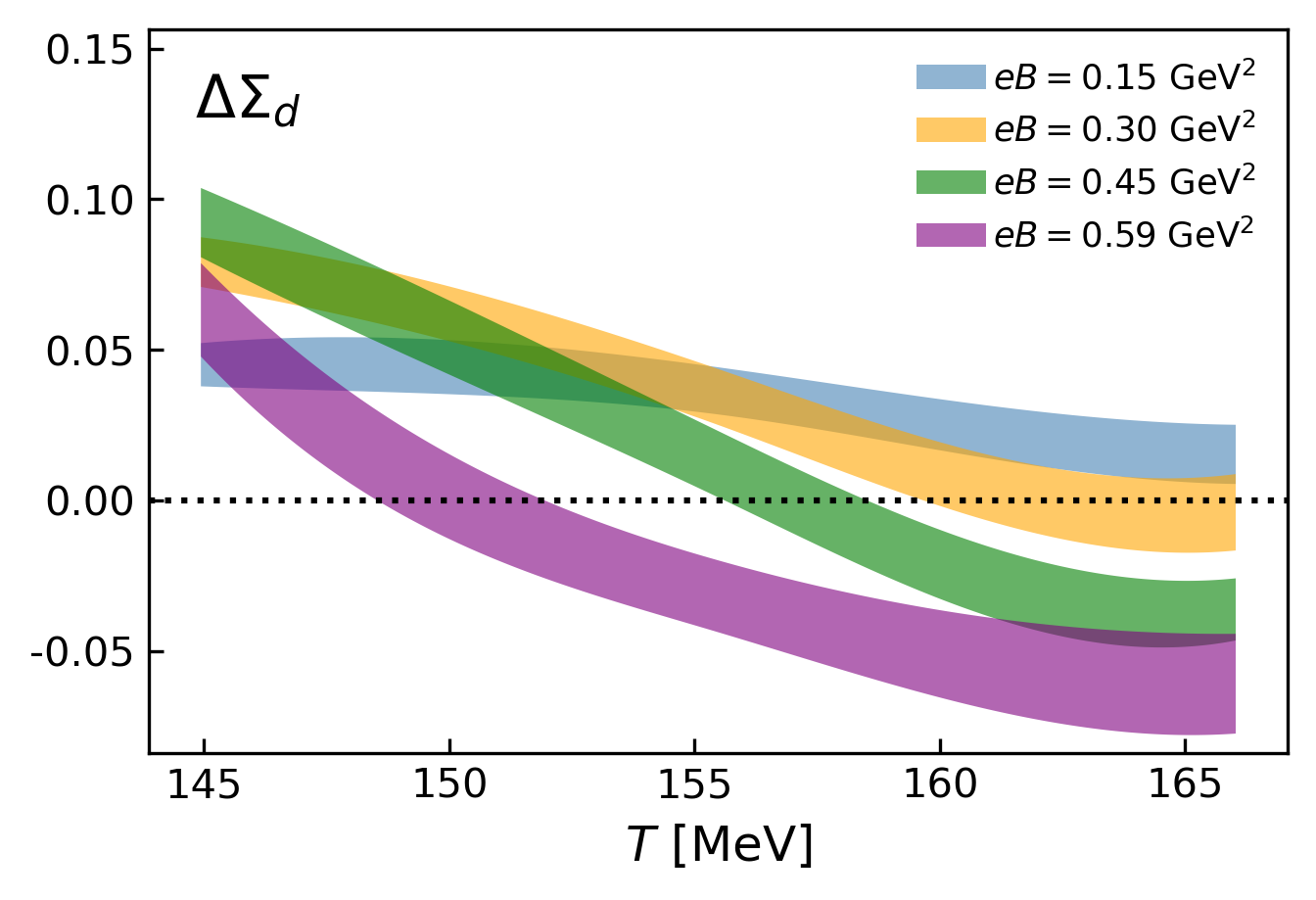}
\caption{ The continuum estimate of the change of the renormalized 
chiral condensates $\Delta\Sigma_{u}$ (top) and $\Delta\Sigma_{d}$ (bottom) as a function of temperature
$T$ at fixed magnetic field strength $eB$. }
\label{pbpq_vs_T}
\end{figure}

We use a 2-dimensional B-spline interpolation in $T-eB$ plane to fit our lattice data and obtain the intermediate values of lattice quantities for each lattice size. The interpolation algorithm inherently determines the number of knots and their position while the smoothing factor is provided to control the smoothing \cite{dierckx1981algorithm}. A similar approach of using a 2-dimension interpolation in $eB - T$ plane was also used in \cite{Bali:2011qj,Ding:2023bft}. The error bands are obtained by bootstrap sampling the data and performing the spline interpolation on each sample, choosing the median as our central estimate and 68.27\% quantile as the error estimate. 

\autoref{fig:interpolation} shows the interpolation estimation of the screening masses of neutral pseudoscalar mesons as functions of the magnetic field $eB$ at fixed temperatures, based on lattice QCD calculations. Each panel corresponds to a specific lattice size: $32^3\times 8$ for $\pi^0$ (top), $48^3\times 12 $ for $K^0$ (middle), $64^3\times 16$ for $\eta^0_{s\bar{s}}$ (bottom). The data points represent the lattice simulation results, while the shaded bands show the interpolated values obtained through a 2-dimensional spline fit in the $T-eB$ plane. The interpolation captures the variation of screening masses across different temperatures and magnetic field strengths, with the shaded bands reflecting the interpolated trends.

The lattice action employed introduces discretization effects at $\mathcal{O}(a^2)$. To mitigate these lattice artifacts and obtain continuum estimates of physical observables, we perform an expansion of the lattice observables in powers of $1/N_\tau^2$. The continuum limit is approached by extrapolating the results using fits to the $T-eB$ interpolated data. For lattices with $N_\tau = 12$ and $16$, we apply a linear ansatz of the form:  
\begin{eqnarray}
    O(T,eB,N_\tau) = O^{\text{lin}}(T,eB) + \frac{b}{N_\tau^2},
    \label{linear_ansatz}
\end{eqnarray}
where $b$ is a fit parameter and for lattices with $N_\tau = 8, 12, \text{ and } 16$, we employ a quadratic ansatz:  
\begin{eqnarray}
    O(T,eB,N_\tau) = O^{\text{quad}}(T,eB) + \frac{c}{N_\tau^2} + \frac{d}{N_\tau^4},
    \label{quad_ansatz}
\end{eqnarray}
where $c$ and $d$ are fit parameters. The final continuum estimate is obtained by bootstrapping the averaged results from the two ansatz. The central value is determined by the median of the averaged results, while the associated uncertainty is quantified using the 68.27\% quantile of the pooled distribution.

\autoref{sigma_estimate} shows samples of the continuum estimate of the difference for light chiral condensate $\Sigma_{ud}$ at fixed temperatures $T$ and magnetic field strengths $eB$. Results are shown for three different temperatures: T=145.0 MeV (top), T=157.0 MeV (middle), and at T=166.0 MeV (bottom). The data points are obtained through interpolation of the lattice data, and the shaded bands depict the continuum estimate error. 

Similarly, \autoref{estimate} presents samples of the continuum estimate of screening masses at fixed temperatures $T$ and magnetic field strengths $eB$. Results are shown for three neutral pseudoscalar mesons at representative temperatures: $\pi^0$ at T=145.0 MeV (top), $K^0$ at T=157.0 MeV (middle), and  $\eta^0_{s\bar{s}}$ at T=166.0 MeV (bottom). The data points are obtained through interpolation of the lattice data, and the shaded bands depict the continuum estimate error.

\begin{table*}[]
\begin{tabular}{|c|c|c|c|c|c|c|c|c|c|c|}                                                     
\hline          
& \multicolumn{2}{c|}{$\beta = 6.315$} &\multicolumn{2}{c|}{$\beta = 6.354$} &\multicolumn{2}{c|}{$\beta = 6.390$} &\multicolumn{2}{c|}{$\beta = 6.423$} &\multicolumn{2}{c|}{$\beta = 6.445$}     \\       
& \multicolumn{2}{c|}{$am_s=0.0759$} & \multicolumn{2}{c|}{$am_s=0.0728$} & \multicolumn{2}{c|}{$am_s=0.0694$} & \multicolumn{2}{c|}{$am_s=0.0670$} & \multicolumn{2}{c|}{$am_s=0.0652$} \\
& \multicolumn{2}{c|}{$T = 144.95$ MeV} &\multicolumn{2}{c|}{$T = 151.00$ MeV} &\multicolumn{2}{c|}{$T = 156.78$ MeV}&\multicolumn{2}{c|}{$T = 162.25$ MeV} &\multicolumn{2}{c|}{$T = 165.98$ MeV}      \\ 
\hline
Nb& $\Delta\Sigma$ &$M^{scr}$& $\Delta\Sigma$ &$M^{scr}$& $\Delta\Sigma$ &$M^{scr}$& $\Delta\Sigma$ &$M^{scr}$& $\Delta\Sigma$ &$M^{scr}$\\
\hline          
 1  & 54472 &   5724 & 53047 &  4714 & 61861  &   5785  & 24859  &   6913  &  23431  &   7137 \\
 2  & 53230 &   5449 & 52390 &  4899 & 54868  &   5994  & 22398  &   7257  &  18697  &   7338 \\
 3  & 49959 &   4765 & 59699 &  4602 & 45294  &   5860  & 25189  &   6566  &  21551  &   7298 \\
 4  & 59152 &   4720 & 59367 &  4421 & 50639  &   5739  & 24787  &   6371  &  20501  &   7170 \\
 6  & 60859 &   8042 & 58573 &  9108 & 45874  &   9773  & 29553  &   9279  &  21503  &   9343 \\
 12 & 59762 &   8014 & 59604 &  8043 & 43767  &   8316  & 34413  &   8530  &  25733  &   6983 \\
 16 & 48174 &   8103 & 50114 &  8262 & 44571  &   7599  & 42203  &   9030  &  27661  &   7167 \\  
 24 & 48781 &   8815 & 41861 &  7382 & 38098  &   8468  & 39571  &   9577  &  47784  &   7743 \\
\hline  
    \end{tabular}
    \caption{Configuration statistics measured for chiral condensate $\Delta\Sigma$ and screening mass $M^{scr}$ analysis for lattice size $32^3\times 8$.}  
    \label{tab:328_new}                                                                      
\end{table*}

\begin{table*}[]
\begin{tabular}{|c|c|c|c|c|c|c|c|c|c|c|}
\hline
& \multicolumn{2}{c|}{$\beta = 6.712$} &\multicolumn{2}{c|}{$\beta = 6.754$} &\multicolumn{2}{c|}{$\beta = 6.794$} &\multicolumn{2}{c|}{$\beta = 6.825$} &\multicolumn{2}{c|}{$\beta = 6.850$}     \\       
& \multicolumn{2}{c|}{$am_s=0.0490$} & \multicolumn{2}{c|}{$am_s=0.0468$} & \multicolumn{2}{c|}{$am_s=0.0450$} & \multicolumn{2}{c|}{$am_s=0.0436$} & \multicolumn{2}{c|}{$am_s=0.0424$} \\
& \multicolumn{2}{c|}{$T = 144.97$ MeV} &\multicolumn{2}{c|}{$T = 151.09$ MeV} &\multicolumn{2}{c|}{$T = 157.13$ MeV}&\multicolumn{2}{c|}{$T = 161.94$ MeV} &\multicolumn{2}{c|}{$T = 165.91$ MeV}      \\ 
\hline
Nb& $\Delta\Sigma$ &$M^{scr}$& $\Delta\Sigma$ &$M^{scr}$& $\Delta\Sigma$ &$M^{scr}$& $\Delta\Sigma$ &$M^{scr}$& $\Delta\Sigma$ &$M^{scr}$\\
\hline
 1  & 10883 & 3157  & 9075 & 3132  & 8434 &3347  & 6154 & 6154   & 6342 &  6342  \\           
 2  & 11182 & 3157  & 8623 & 2964  & 8620 &3553  & 6170 & 6170   & 5559 &  5559  \\        
 3  & 7180  & 3157  & 8589 & 2973  & 7675 &3197  & 6785 & 6786   & 4470 &  4468  \\        
 4  & 8632  & 3157  & 8974 & 3115  & 6660 &3309  & 7318 & 7318   & 4779 &  4778  \\        
 6  & 6469  & 3157  & 9157 & 3381  & 8386 &3120  & 7674 & 7674   & 3829 &  3829  \\           
 12 & 7146  &  3880 & 5990 &  3519 & 6028 & 3125 & 4732 &  3005  & 5321 &   3005 \\         
 16 & 7277  &  3840 & 5414 &  3114 & 6345 & 3343 & 4386 &  2942  & 5919 &   3005 \\           
 24 & 6093  &  3032 & 5412 &  3087 & 6518 & 3205 & 4263 &  3005  & 5863 &   3005 \\
\hline
    \end{tabular}
    \caption{Configuration statistics measured for chiral condensate $\Delta\Sigma$ and screening mass $M^{scr}$ analysis for lattice size $48^3\times 12$.}
    \label{tab:4812_new}
\end{table*}

\begin{table*}[]
\begin{tabular}{|c|c|c|c|c|c|c|c|c|c|c|}
\hline
& \multicolumn{2}{c|}{$\beta = 7.010$} &\multicolumn{2}{c|}{$\beta = 7.054$} &\multicolumn{2}{c|}{$\beta = 7.095$} &\multicolumn{2}{c|}{$\beta = 7.130$} &\multicolumn{2}{c|}{$\beta = 7.156$}     \\       
& \multicolumn{2}{c|}{$am_s=0.0357$} & \multicolumn{2}{c|}{$am_s=0.0348$} & \multicolumn{2}{c|}{$am_s=0.0334$} & \multicolumn{2}{c|}{$am_s=0.0322$} & \multicolumn{2}{c|}{$am_s=0.0314$} \\
& \multicolumn{2}{c|}{$T = 144.94$ MeV} &\multicolumn{2}{c|}{$T = 151.04$ MeV} &\multicolumn{2}{c|}{$T = 156.92$ MeV}&\multicolumn{2}{c|}{$T = 162.10$ MeV} &\multicolumn{2}{c|}{$T = 166.03$ MeV}      \\ 
\hline
Nb& $\Delta\Sigma$ &$M^{scr}$& $\Delta\Sigma$ &$M^{scr}$& $\Delta\Sigma$ &$M^{scr}$& $\Delta\Sigma$ &$M^{scr}$& $\Delta\Sigma$ &$M^{scr}$\\
\hline
 1  & 9274 & 3510 & 4899 & 3370 & 4807 & 4807 & 4505 & 3566  & 4604 & 3968    \\
 2  & 6988 & 3510 & 4624 & 3449 & 4707 & 4707 & 5594 & 3976  & 5739 & 3538    \\
 3  & 8043 & 3510 & 5237 & 3400 & 5467 & 3537 & 5712 & 3557  & 3931 & 3664    \\
 4  & 7872 & 3510 & 5786 & 3510 & 5707 & 4212 & 5788 & 3843  & 5183 & 2948    \\
 6  & 8491 & 3510 & 5814 & 3323 & 5130 & 3240 & 5871 & 3904  & 4349 & 3464    \\
 12 & 6826 &  4010& 5777 &  3424& 5618 &  3149& 4164 &  4010 & 4256 &  3624   \\
 16 & 5426 &  3822& 5857 &  2731& 5452 &  2060& 5854 &  3863 & 5174 &  2010   \\
 24 & 3483 &  2199& 3833 &  2501& 3573 &  2033& 3941 &  3572 & 5905 &  2256   \\
\hline
    \end{tabular}
    \caption{Configuration statistics measured for chiral condensate $\Delta\Sigma$ and screening mass $M^{scr}$ analysis for lattice size $64^3\times 16$.}
    \label{tab:6416_new}
\end{table*}

\section{Separate behavior of the renormalized chiral condensates for $u$ and $d$ quarks}
\label{udquark}

\def\sectionautorefname{Appendix}
In this appendix, we examine separately the behavior of the change in the renormalized chiral condensate for $u$ and $d$ quarks as functions of $eB$ and temperature. We define
\begin{eqnarray}
    \Delta\Sigma_{f}(B,T) &= \frac{m_f}{M_\pi^2f_{\pi}^2}& \left\{\pbp_f(B,T) - \pbp_f(0,T)\right\}\nonumber\\
\end{eqnarray}
for $f=u,d$, such that $\Delta\Sigma_{ud}=\Delta\Sigma_{u}+\Delta\Sigma_{d}$.

In \autoref{pbpq_vs_eb}, we show the change of renormalized chiral condensate a function of the magnetic field strength $eB$ at several fixed temperatures, separately for $\Delta\Sigma_u$ (top) and $\Delta\Sigma_d$ (bottom). Although the overall behavior of $\Delta\Sigma_{u}$ and $\Delta\Sigma_{d}$ is qualitatively similar to each other and that of $\Delta\Sigma_{ud}$ in \autoref{pbp_vs_eb}, a notable difference arises in their numerical magnitudes. Specifically, 
$\Delta\Sigma_{u}$ exhibits larger values compared to $\Delta\Sigma_{d}$  across all temperatures studied.

In \autoref{pbpq_vs_T}, we plot the change of renormalized chiral condensate a function of the temperature $T$ for fixed magnetic field strength $eB$, again separately for $\Delta\Sigma_u$ (top) and $\Delta\Sigma_d$ (bottom). Here, we also observe a qualitatively similar behavior between $\Delta\Sigma_{u}$, $\Delta\Sigma_{d}$ and $\Delta\Sigma_{ud}$, particularly regarding the characteristic crossing behavior of curves corresponding to constant $eB$. However, the crossing points occur at slightly different temperatures, with $\Delta\Sigma_{d}$ having the lowest crossing temperature, followed by $\Delta\Sigma_{ud}$ and finally $\Delta\Sigma_{u}$.

\section{Statistics}
\label{statistics}
In~\autoref{tab:328_new},~\autoref{tab:4812_new} and \autoref{tab:6416_new} the parameters and number of configurations used for the analysis of chiral condensate as well as screening mass in the lattice simulations are listed for lattice sizes of $32^3\times8$, $48^3\times12$ and $64^3\times16$, respectively. All configurations used are separated by ten time units.

\end{document}